\input harvmac
%%% Figures
%
\ifx\epsfbox\UnDeFiNeD\message{(NO epsf.tex, FIGURES WILL BE
IGNORED)}
\def\figin#1{\vskip2in}% blank space instead
\else\message{(FIGURES WILL BE INCLUDED)}\def\figin#1{#1}\fi
\def\ifig#1#2#3{\xdef#1{fig.~\the\figno}
\goodbreak\midinsert\figin{\centerl ine{#3}}%
\smallskip\centerline{\vbox{\baselineskip12pt
\advance\hsize by -1truein\noindent\footnotefont{\bf
Fig.~\the\figno:} #2}}
\bigskip\endinsert\global\advance\figno by1}
%%%%%%%%%%%%%%%%%%%%%%%%%%%%%%%%%%%%%%%%%%%%%%%%%%%%%%%%%%%%%%%%

\def\footnotefont{\tenpoint}

\newwrite\ffile\global\newcount\figno \global\figno=1
\def\fig{fig.~\the\figno\nfig}
\def\nfig#1{\xdef#1{fig.~\the\figno}%
\writedef{#1\leftbracket fig.\noexpand~\the\figno}%
\ifnum\figno=1\immediate\openout\ffile=figs.tmp\fi\chardef\wfile=\ffile%
\immediate\write\ffile{\noexpand\medskip\noexpand\item{Fig.\
\the\figno. }
\reflabeL{#1\hskip.55in}\pctsign}\global\advance\figno
by1\findarg}
%\ionew

\def\dal{{\dot \alpha}}
\def\dbet{{\dot \beta}}
\def\bmu{{\bar \mu}}

\overfullrule=0pt
\parindent 25pt
\tolerance=10000
%\sequentialequations
%\draftmode

\def\NSNS{{$NS\otimes NS$}}
\def\RR{{$R\otimes R$}}

\def\det{\hbox{\rm det}}

\def\tr{{\rm tr}}

\def\G(#1){\Gamma(#1)}

\def\half{{\textstyle {1 \over 2}}}

\def\(#1#2){(\zeta_#1\cdot  \zeta_#2)}

\def\Mp{{M'}}
\def\tPhi{{\tilde \Phi}}

\def\C|#1{{\cal #1}}
\def\calA{{\cal A}}
\def\calW{{\cal W}}

\def\calN{{\cal N}}

\def\dalp{\dot \alpha}

%%%%%%%References%%%%%%%
\def\lr{\lref}
\lr\rfIengoZhu{R.~Iengo and C.~Zhu,
    {\it Explicit modular invariant two-loop superstring amplitude
    relevant for  R**4,}  JHEP {\bf 06}  (1999) 011, hep-th/9905050.}
\lr\veneziano{G.~Veneziano,
``Construction Of A Crossing - Symmetric, Regge Behaved Amplitude For
Linearly Rising Trajectories,''
Nuovo.\ Cim.\ {\bf A57} (1968) 190.}
\lr\amati{V.~Alessandrini, D.~Amati and B.~Morel, Nuovo.\ Cim.\ {\bf 7A}
(1971) 797.}
\lr\grossmende{D.J.~Gross and P.F.~Mende,
``String Theory Beyond The Planck Scale,''
Nucl.\ Phys.\ {\bf B303} (1988) 407.}
\lr\gutbound{M.~Gutperle,
``Multiboundary effects in Dirichlet string theory,''
Nucl.\ Phys.\ {\bf B444} (1995) 487
[hep-th/9502106].}
\lr\mli{M.~Li, ``Dirichlet strings,''
Nucl.\ Phys.\ {\bf B420} (1994) 339
[hep-th/9307122].}

\lr\greentwob{M.B.~Green,
``Pointlike states for type 2b superstrings,''
Phys.\ Lett.\ {\bf B329} (1994) 435
[hep-th/9403040].}
\lr\greenguteffects{M.B.~Green and M.~Gutperle,
``Effects of D-instantons,''
Nucl.\ Phys.\ {\bf B498} (1997) 195
[hep-th/9701093].}
\lr\daileighpolch{J.~Dai, R.G.~Leigh and J.~Polchinski,
``New Connections Between String Theories,''
Mod.\ Phys.\ Lett.\ {\bf A4} (1989) 2073.}
\lr\polchinski{J.~Polchinski,
``Dirichlet-Branes and Ramond-Ramond Charges,''
Phys.\ Rev.\ Lett.\ {\bf 75} (1995) 4724
[hep-th/9510017].}
\lr\polchdinst{J.~Polchinski,
``Combinatorics of boundaries in string theory,''
Phys.\ Rev.\ {\bf D50} (1994) 6041
[hep-th/9407031].}
\lr\douglasa{M.R.~Douglas,
``Gauge Fields and D-branes,''
J.\ Geom.\ Phys.\ {\bf 28} (1998) 255
[hep-th/9604198].}
\lr\bachaspioline{C.~Bachas and B. Pioline, ``High energy scattering on
distant branes'', [hep-th/9909171].}

\lr\mbgsethi{M.B.~Green and S.~Sethi,
``Supersymmetry constraints on type IIB supergravity,''
Phys.\ Rev.\ {\bf D59} (1999) 046006
[hep-th/9808061].}
\lr\klebth{I.R.~Klebanov and L.~Thorlacius,
``The Size of p-Branes,''
Phys.\ Lett.\ {\bf B371} (1996) 51
[hep-th/9510200].}
\lr\myers{ M.R.~Garousi and R.C.~Myers,
``Superstring Scattering from D-Branes,''
Nucl.\ Phys.\ {\bf B475} (1996) 193
[hep-th/9603194].}
\lr\bainbachasgreen{C.P.~Bachas, P.~Bain and M.B.~Green,
``Curvature terms in D-brane actions and their M-theory origin,''
JHEP {\bf 05} (1999) 011
[hep-th/9903210].}
\lr\greenvanhove{M.B.~Green and P.~Vanhove,
``D-instantons, strings and M-theory,''
Phys.\ Lett.\ {\bf B408} (1997) 122
[hep-th/9704145].}
\lr\greenharveymoore{M.B.~Green, J.A.~Harvey and G.~Moore,
``I-brane inflow and anomalous couplings on D-branes,''
Class.\ Quant.\ Grav.\ {\bf 14} (1997) 47
[hep-th/9605033].}
\lr\greenguta{M.B.~Green and M.~Gutperle,
``Comments on Three-Branes,''
Phys.\ Lett.\ {\bf B377} (1996) 28
[hep-th/9602077].}
\lr\tseytlina{A.A.~Tseytlin,
``Self-duality of Born-Infeld action and Dirichlet 3-brane of type IIB
superstring theory,''
Nucl.\ Phys.\ {\bf B469} (1996) 51
[hep-th/9602064].}
\lr\zanon{A.~De Giovanni, A.~Santambrogio and D.~Zanon,
``${\alpha'}^4$ corrections to the N = 2 supersymmetric Born-Infeld
action,''
[hep-th/9907214].}
\lr\shmakova{M.~Shmakova,
``One-loop corrections to the D3 brane action,''
[hep-th/9906239].}
\lr\fradkintseytlin{E.S.~Fradkin and A.A.~Tseytlin,
``Nonlinear Electrodynamics From Quantized Strings,''
Phys.\ Lett.\ {\bf B163} (1985) 123.}
\lr\doreythree{N.~Dorey, T.J.~Hollowood, V.V.~Khoze, M.P.~Mattis and
S.~Vandoren,
``Multi-instanton calculus and the AdS/CFT correspondence in N = 4
superconformal field theory,''
Nucl.\ Phys.\ {\bf B552} (1999) 88
[hep-th/9901128].}
\lr\witteninst{E.~Witten,
``Sigma models and the ADHM construction of instantons,''
J.\ Geom.\ Phys.\ {\bf 15} (1995) 215
[hep-th/9410052]. }

\lr\wessbagger{J.~Bagger and J.~Wess,
``Supersymmetry And Supergravity'', Princeton University Press.}
\lr\susybia{M.~Aganagic, C.~Popescu and J.H.~Schwarz,
``D-brane actions with local kappa symmetry,''
Phys.\ Lett.\ {\bf B393} (1997) 311
[hep-th/9610249].}
\lr\susybib{E.~Bergshoeff and P.K.~Townsend,
``Super D-branes,''
Nucl.\ Phys.\ {\bf B490} (1997) 145
[hep-th/9611173].}
\lr\susybic{M.~Cederwall, A.~von Gussich, B.E.~Nilsson and A.~Westerberg,
``The Dirichlet super-three-brane in ten-dimensional type IIB
supergravity,''
Nucl.\ Phys.\ {\bf B490} (1997) 163
[hep-th/9610148].}
\lr\ooguristudents{Y-K. E. Cheung and Z.  Yin, '' Anomalies, branes and
currents,''   Nucl. Phys. {\bf B517} (1998) 69
[hep-th/9710206].}
\lr\lercheetal{W.~Lerche and S.~Stieberger,
``Prepotential, mirror map and F-theory on K3,''
Adv.\ Theor.\ Math.\ Phys.\ {\bf 2} (1998) 1105
[hep-th/9804176];
``On the anomalous and global interactions of Kodaira 7-planes,''
[hep-th/9903232].}
\lr\gutperlea{M.~Gutperle,
``A note on heterotic/type I' duality and D0 brane quantum mechanics,''
JHEP {\bf 05} (1999) 007
[hep-th/9903010]; ``Heterotic/type I duality, D-instantons and a N = 2
AdS/CFT  correspondence,''
Phys.\ Rev.\ {\bf D60} (1999) 126001
[hep-th/9905173].}
\lr\review{O.~Aharony, S.S.~Gubser, J.~Maldacena, H.~Ooguri and Y.~Oz,
``Large N field theories, string theory and gravity,''
[hep-th/9905111].}
\lr\malda{J.~Maldacena,
``The large-N limit of superconformal field theories and supergravity,''
Adv.\ Theor.\ Math.\ Phys.\ {\bf 2} (1998) 231
[hep-th/9711200].}
\lr\garmyers{M.R.~Garousi and R.C.~Myers,
``World-volume interactions on D-branes'',  Nucl.Phys. {\bf B542} (1999)
73-88 [hep-th/9809100].}
\lr\hashkleb{A. Hashimoto, I. R. Klebanov, `` Decay of Excited
D-branes'',   hep-th/9604065 Phys. Lett. {\bf B381} (1996) 437-445  [hep-th/9604065];
``Scattering of Strings from D-branes'',
Nucl. Phys. Proc. Suppl. {\bf 55B} (1997) 118-133 [hep-th/9611214]. }
\lr\dixonkapluis{L.~J.~Dixon, V.~Kaplunovsky and J.~Louis,
``Moduli dependence of string loop corrections to gauge coupling constants,''
Nucl.\ Phys.\  {\bf B355} (1991) 649.}
\lr\harveymoore{  J.~A.~Harvey and  G.~Moore, ``Algebras, BPS States, and
Strings'', Nucl.\ Phys.\  {\bf B463} (1996) 315. [hep-th/9510182] }
\lr\bachaskiritsisvanhove{C.~Bachas, C.~Fabre, E.~Kiritsis, N.A.~Obers and
P.~Vanhove, ``Heterotic / type I duality and D-brane instantons'', Nucl.\
Phys.\  {\bf B509} (1998) 33 [hep-th/9707126].}
\lr\kiritsisobers{E.~Kiritsis and  N.A.~Obers, ``Heterotic/Type-I Duality
in $D<10$ Dimensions, Threshold Corrections and D-Instantons'', JHEP {\bf
10} (1997) 004 [hep-th/9709058].}
\lr\nekrasovschwarz{N.~Nekrasov and  A.~Schwarz, ``Instantons on
noncommutative $R^4$, and (2,0) superconformal six dimensional theory'',
Commun.Math.Phys. {\bf 198} (1998) 689 [hep-th/9802068].}
\lr\greengutkwon{ M.B.~Green, M.~Gutperle
and  Hwang-hyun~Kwon, `` Sixteen-fermion and related terms in M-theory on
$T^2$'',  Phys.Lett. {\bf B421} (1998) 149 [hep-th/9710151]. }
\lr\howest{P.~S.~Howe and P.~C.~West,
``The Complete N=2, D = 10 Supergravity,''
Nucl.\ Phys.\  {\bf B238} (1984) 181.}
\lr\greenschwarza{M.~B.~Green and J.~H.~Schwarz,
``Supersymmetrical Dual String Theory. 2. Vertices And Trees,''
Nucl.\ Phys.\  {\bf B198} (1982) 252.}
\lr\dzerodfour{M.~R.~Douglas, D.~Kabat, P.~Pouliot and S.~H.~Shenker,
``D-branes and short distances in string theory,''
Nucl.\ Phys.\  {\bf B485} (1997) 85
[hep-th/9608024].}
\lr\aharonyseib{O.~Aharony, M.~Berkooz and N.~Seiberg,
``Light-cone description of (2,0) superconformal theories in six  dimensions,''
Adv.\ Theor.\ Math.\ Phys.\  {\bf 2} (1998) 119
[hep-th/9712117].}
\lr\wittenmoduli{E.~Witten,
``Sigma models and the ADHM construction of instantons,''
J.\ Geom.\ Phys.\  {\bf 15} (1995) 215
[hep-th/9410052].}
\lr\eisen{L.P.~Eisenhart, Riemannian Geometry, Princeton University
Press, 1926.}
\lr\koba{S.~Kobayashi and K.~Nomizu, Foundations of Differential
Geometry, J. Wiley, New York 1969.}
\lr\mbgitaly{M.~Bianchi, M.~B.~Green, S.~Kovacs and G.~Rossi,
``Instantons in supersymmetric Yang-Mills and D-instantons in IIB  superstring theory,''
JHEP {\bf 9808} (1998) 013
[hep-th/9807033].}
\lr\greensethi{M.~B.~Green and S.~Sethi,
``Supersymmetry constraints on type IIB supergravity,''
Phys.\ Rev.\  {\bf D59} (1999) 046006
[hep-th/9808061].}
\lr\gsw{M.~B.~Green, J.~H.~Schwarz and E.~Witten,
``Superstring Theory. Vol. 1: Introduction,''
{  Cambridge  University Press ( 1987). }}
\lr\townsendsuper{P.~Howe, K.~S.~Stelle and P.~K.~Townsend,
``Supercurrents,''
Nucl.\ Phys.\  {\bf B192}, 332 (1981).}
\lr\deroo{M.~de Roo,
``Matter Coupling In N=4 Supergravity,''
Nucl.\ Phys.\  {\bf B255} (1985) 515.}
\lr\bergsh{E.~Bergshoeff, I.~G.~Koh and E.~Sezgin,
``Coupling Of Yang-Mills To N=4, D = 4 Supergravity,''
Phys.\ Lett.\  {\bf B155} (1985) 71.}

\lr\seibergwitten{N.~Seiberg and E.~Witten,
``String theory and noncommutative geometry,''
JHEP {\bf 9909} (1999) 032
[hep-th/9908142].}

\lr\convens{M.~R.~Douglas,
``Branes within branes,''
[hep-th/9512077].}
\lr\doreytwo{V.~V.~Khoze, M.~P.~Mattis and M.~J.~Slater,
``The instanton hunter's guide to supersymmetric SU(N) gauge theory,''
Nucl.\ Phys.\  {\bf B536} (1998) 69
[hep-th/9804009].}
\lr\borisetala{I.~Antoniadis, B.~Pioline and T.~R.~Taylor,
``Calculable $e^{-1/\lambda}$ effects,''
Nucl.\ Phys.\  {\bf B512} (1998) 61
[hep-th/9707222].}
\lr\borisetalb{A.~Gregori, E.~Kiritsis, C.~Kounnas, N.~A.~Obers, P.~M.~Petropoulos and B.~Pioline,
``$R^2$ corrections and non-perturbative dualities of N = 4 string ground  states,''
Nucl.\ Phys.\  {\bf B510} (1998) 423
[hep-th/9708062].}

\lr\nakajima{H.~Nakajima, ``Resolutions of moduli spaces of ideal
instantons on $R^4$', in 'Topology, Geometry and Field Theory', (World
Scientific,1994).}

\lr\tseytlind{O.~D.~Andreev and A.~A.~Tseytlin,
``Partition Function Representation For The Open
 Superstring Effective Action: Cancellation Of Mobius Infinities
 And Derivative Corrections To
Born-Infeld Lagrangian,''
Nucl.\ Phys.\  {\bf B311} (1988) 205.}
\lr\tseytline{A.~A.~Tseytlin,
``Born-Infeld action, supersymmetry and string theory,''
[hep-th/9908105].}
%%%%%%%%%%%%%%%%%%%%%%%%%%%%%%%%%%%%%%%%%%%%%%%%%%%%%%%%%%%%%%%%%%%
%%%%%%%%% title and abstract
%%%%%%%%%%%%%%%%%%%%%%%%%%%%%%%%%%%%%%%%%%%%%%%%%%%%%%%%%%%%%%%%%%%
\noblackbox
\baselineskip 14pt plus 2pt minus 2pt
\Title{\vbox{\baselineskip12pt \hbox{hep-th/0002011}
\hbox{DAMTP-2000-6} \hbox{HUTP-00/A001}  }}
{\vbox{ \centerline{D-instanton induced interactions on a D3-brane} }}

\centerline{Michael B. Green}
\medskip
\centerline{DAMTP, Wilberforce Road, Cambridge CB3 0WA, UK}
\centerline{\tt m.b.green@damtp.cam.ac.uk}
\medskip
\centerline{Michael Gutperle}
\medskip
\centerline{Physics Department, Harvard University, Cambridge, MA 02138,
USA}
\centerline{\tt gutperle@riemann.harvard.edu}
\bigskip

%% abstract
\medskip
\centerline{{\bf Abstract}}

Non-perturbative features of the derivative expansion of the
 effective action of a single
D3-brane  are obtained by considering scattering amplitudes of
open and closed strings. This motivates  expressions for the
coupling constant dependence of world-volume interactions of the
form $(\partial F)^4$ (where $F$ is the Born--Infeld field
strength), $(\partial^2\varphi)^4$ (where $\varphi$ are the normal
coordinates of the D3-brane) and other interactions related by
$\calN=4$ supersymmetry.
  These include terms that transform with non-trivial modular weight under
Montonen--Olive duality.  The leading D-instanton  contributions that
enter into these effective interactions are also shown to follow
from an explicit stringy  construction of the
 moduli space action for the
D-instanton/D3-brane system in the presence of D3-brane
open-string sources (but in the absence of a  background antisymmetric tensor
potential). Extending this action to
include closed-string sources leads to a unified description of
non-perturbative terms in the effective action of the form
$($embedding curvature$)^2$ together with open-string interactions
that describe contributions of the second fundamental form.

%%%%%%%%%%%%%%%%%%%%%%%%%%%%%%%%%%%%%%%%%%%%%%%%%%%%%%%%%%%%%%%%%%%
\noblackbox
\baselineskip 14pt plus 2pt minus 2pt

\Date{January 2000}

%%%%%%%%%%%%%%%%%%%%%%%%%%%%%%%%%%%%%%%%%%%%%%%%%%%%%%%%%%%%%%%%%%%

\newsec{Introduction}

This paper concerns properties of scattering amplitudes of open and closed strings
 on a D-brane and their implications for  the low energy effective world-volume action.
  We will be particularly interested in
 non-perturbative effects associated with the presence
 of D-instantons (or D(-1)-branes) which
 are essential in ensuring the $SL(2,Z)$  invariance of  type IIB string theory.
The scattering of open string states  describes  the interactions
of both the Born--Infeld world-volume gauge field and of the
scalar world-volume fields,  while interactions between closed and
open strings describe gravitational effects induced  on the brane
\refs{\klebth,\myers}.  These curvature-dependent effects result
both from the embedding of the D-brane in a geometrically
non-trivial target space as well as from the non-trivial intrinsic
geometry of the D-brane. The long wavelength dependence of such
effects is  encoded in the derivative expansion of the effective
world-volume action of the D-brane.  Of course, the
Dirac--Born--Infeld (DBI) part of the D-brane action already
contains an infinite number of higher derivative terms since it
can be expanded in an infinite power series in the Born--Infeld
field strength, $F$.
 However, this only accounts for constant
 $F$'s while we will be concerned with terms that depend on the first
derivative of $F$  \refs{\tseytlind,\tseytline}.
These arise as natural partners of terms in
the world-volume action such as those of the form  $R^2$, which
denotes the sum of a number of $($curvature$)^2$ terms
\refs{\bainbachasgreen}. These include  both normal and tangential
components of the pull-back of the  curvature together with the
contributions that come from the second fundamental form (which
depends on the scalar world-volume fields) that enter for
non-geodesic embeddings.

We will focus particularly on properties of the
D3-brane,
which is a system of obvious intrinsic interest in the context of
four-dimensional field theory.  The requirement that the
 equations of motion derived from the action be
invariant under
$SL(2,Z)$ duality transformations provides very strong constraints on the
possible
structure of higher dimensional  terms  in the world-volume action, just
as in the case of the bulk effective
string action \refs{\greenguteffects,\greenvanhove}.
As we will see, this motivates a non-perturbative
expression   for low-lying terms in the derivative expansion of the action
that includes an exact description of the effects of D-instantons
 that are localized on the D3-brane.

In section 2 we will review the amplitude that describes the scattering of
four massless
open-string states on a D3-brane.  The low energy expansion of the
well-known expression for the tree amplitude  leads to
contributions in the effective action that are
 of order ${\alpha'}^4$ relative to the classical
Yang--Mills amplitude.\foot{We will always count powers of $\alpha'$
relative to the  $F^2$ term in this paper.}  These terms include one of the
form $(\partial F)^4$ and one of the form $(\partial^2 \varphi)^4$, where
the six
scalar fields $\varphi^a$ ($a=1,\dots, 6$)
 describe the transverse oscillations of the D3-brane (and the
contractions between
 the fields and the derivatives will be specified later).
It is easy to argue that these interactions also receive corrections at
one string
loop \refs{\shmakova,\zanon} but they are not expected to get corrections
from higher-order perturbative effects.  The  invariance of the
effective action under $SL(2,Z)$ Montonen--Olive duality transformations of the
complex coupling constant
will also be discussed in section 2.  This requires that the
dependence on the complex  coupling constant (the type IIB complex
scalar field, $\tau$) enters the interaction via a modular invariant prefactor,
$h(\tau,\bar\tau)$.
We will argue that the known perturbative
contributions to the
higher derivative interactions are consistent with  $h(\tau,\bar\tau)$
having the form,
\eqn\modfun{h(\tau,\bar\tau)= \ln|\tau_2 \eta(\tau)^4|,}
where $\eta$ is the Dedekind function  and $\tau$ is the complex
background scalar field $\tau= \tau_1+i\tau_2 =  C^{(0)}+ ie^{-\phi}$.
The string coupling is $g= e^\phi$ where $\phi$ is the type IIB
dilaton  and $C^{(0)}$ is the Ramond--Ramond (\RR) scalar. The
function $h$ has the weak coupling (large $\tau_2$) expansion,
\eqn\expanh{h(\tau,\bar\tau)= \Big(- {\pi \over 3}{\tau_2} +  \ln\tau_2- 2
  \sum_{N=1}^\infty
\sum_{m|N}{1\over m}\big( e^{2\pi i\ N \tau}+ e^{-2\pi i\ N \bar
\tau }\big)\Big),}
which  contains the expected power-behaved terms that are identified with
 tree-level and one-loop
terms of open-string perturbation theory,
as well as a specific infinite set of D-instanton
corrections which will be discussed in the following sections.
 The function \modfun\  is the same modular function that appears in several
other contexts
\refs{\bainbachasgreen,\greenvanhove,\dixonkapluis,\harveymoore,
\bachaskiritsisvanhove,
\kiritsisobers,\borisetala,\borisetalb}.

In section 3 we will describe   the bosonic and fermionic
collective coordinates of a single D3-brane in the presence of a
D-instanton.  This is a $1/4$-BPS system that preserves eight of
the 32 components of the ten-dimensional type II supersymmetry. As
in \refs{\convens,\dzerodfour,\doreythree}   we will
motivate the description of the collective coordinates by
considering the toroidal compactification of the composite
D$p$/D$(p+4)$ system. We will make particular reference to the
D5/D9 system as a simple way of enumerating the open-string fields
and their interactions.  Upon toroidal compactification combined
with T-duality this reduces to other well-studied systems such as
 the D0/D4 system in which the D0-brane is described by supersymmetric
 quantum mechanics on instanton moduli space \refs{\dzerodfour,\aharonyseib}.
 Another well-studied
 example is the D1/D5 system in which the string is described by a
two-dimensional
 $(4,4)$ supersymmetric sigma model with
 the instanton moduli space as the target manifold \wittenmoduli.
Compactification and T-duality on $T^6$ leads to a description of
the D-instanton/D3-brane system in which the open strings that end
on the D-instanton describe the isolated states corresponding to
collective coordinates rather than to fields.  It is essential to
integrate over these coordinates  in evaluating  scattering
amplitudes. We will emphasize the origin of these collective
coordinates and their interactions from the insertion of open
string vertex operators on a world-sheet that is a disk with a
segment  of `Neumann' boundary (on which the D3-brane boundary
conditions are satisfied) and a segment of `Dirichlet' boundary
(on which the D-instanton boundary conditions are satisfied).
This gives an efficient procedure for describing the eight
components of  unbroken supersymmetry and the twenty-four
components of broken supersymmetry of the system.

In section 4 we will consider the scattering of open-string states
on the D3-brane in the background of a D-instanton to lowest order
in the string coupling.  This involves
the coupling of open-string sources to the Neumann boundary
segments, from which we deduce the moduli space action for the
D-instanton in the presence of a D3-brane source.  This action is
explicitly invariant under the twenty-four non-linearly realized
broken supersymmetries as well as eight linearly realized unbroken
supersymmetries.  The unbroken supersymmetry transformations of
the three-brane fields (the ${\cal N}=4$ Maxwell multiplet) differ
from those of the free field theory by a simple term involving the
collective coordinates of the D$(-1)$/D3 system.

This action defines  a generating function for the leading
perturbative contribution to the correlation functions
of massless open-string states in  a D-instanton background, which will  be
considered in
section 5.   Integration over the collective coordinates gives
the leading order (in powers of the string coupling) D-instanton
contribution to scattering amplitudes of ground-state open strings on the
D3-brane.  These determine terms in the effective world-volume
 action that are   proportional to  the higher-derivative
 interactions,
such as the $(\partial F)^4$ and $(\partial^2 \varphi)^4$, that
arose in the tree-level analysis of section 2.  These terms have a
dependence on the coupling of the form $e^{2\pi i \tau}$,
 which agrees with   the leading $N=1$ instanton term in the
expansion \expanh\ of the
conjectured  exact form of the modular invariant effective action.
This analysis extends  to  the $SL(2,Z)$-invariant interactions of
the form $(\partial F)^2
(\partial^2\varphi)^2$,  $F^+\partial^2 F^- \partial^2 \Lambda
\partial \bar \Lambda$, $(\partial\bar\Lambda)^2(\partial^2\Lambda)^2$ and others,
where $\Lambda^A_\alpha$ and $\bar \Lambda_A^{\dot \alpha}$ are the Weyl fermions of the
${\cal N}=4$ theory.

More generally, as will be evident in section 6, there are
interactions at the same order in $\alpha'$ that transform under $SL(2,Z)$
with non-zero modular weight.  Examples of these are interactions of
the form $(\partial F)^2 (\partial \bar \Lambda)^4$ (which transforms with
holomorphic and
anti-holomorphic weights $(-1,1)$) and  $(\partial
\bar\Lambda)^8$ (which transforms with holomorphic and
anti-holomorphic weights $(-2,2)$).  Correspondingly, the coupling
constant must enter in
prefactors, $h^{(1,-1)}(\tau,\bar\tau)$ and
$h^{(2,-2)}(\tau,\bar\tau)$,
 that are  modular forms  of compensating  weights.  We will
argue that supersymmetry together with $SL(2,Z)$ invariance
requires that  these prefactors  are
given by applying appropriate modular covariant derivatives to the
modular function $h$.  However, this remains a conjecture that should
be justified by a deeper understanding of the constraints imposed by
${\cal N}=4$ supersymmetry.

The open-string calculations of the earlier sections combine naturally
with  closed-string D-instanton effects  that describe the coupling of
bulk gravity to the
world-volume of the D3-brane.  There are two effects of this kind.
One of these, to be described in detail in section 7, arises from the coupling
of closed strings to the D-instanton.  Just as in the bulk theory
\refs{\greenguteffects} the leading perturbative contribution of this kind is
associated with the coupling of a closed-string vertex operator to
disk diagrams with purely Dirichlet boundary conditions and
with fermionic open strings attached.  These fermionic strings describe the
sixteen fermionic moduli of the bulk D-instanton (so that, for example, an
$R^4$ term is generated in the bulk theory
 from the product of four such disks, each with a
graviton and four fermionic open strings attached).  Integration over
the collective coordinates of the D$(-1)$/D3 system
 soaks  up eight of these fermionic moduli.  The remaining eight fermionic
moduli generate  D-instanton contributions to $R^2$ terms which will be
evaluated explicitly.  These
 have previously been obtained by indirect arguments \refs{\bainbachasgreen}.
    We will see   that
 these instanton contributions package together with the
 open-string interactions to give the nonperturbative  generalization of the
 complete  tree-level $R^2$
 term.  This term includes the effects of nongeodesic
 embeddings of the D3-brane in a general target space which involves
 contributions of the   second fundamental form.

 The second effect arises from the  coupling of  the  \NSNS\ antisymmetric
 tensor potential ($B$) to the combined D(-1)/D3 system and
 is described by  attaching  a
closed-string $B$ vertex operator to the disk with
 two boundary twist operators.  This coupling combines with the
analogous coupling of the  Born--Infeld field
in the  usual gauge-invariant combination, $B +
 2\pi \alpha' F$.   With a non-vanishing background
value  for $B$  the instanton give a non-trivial  $\alpha' \to
 0$ decoupling limit, even in the abelian case. It is
 well known  \refs{\nekrasovschwarz} that  for $B\ne 0$
  the  singular abelian instanton of
  ${\cal N}=4$  Maxwell theory is regularized and is equivalent to a
  Fayet-Illipoulos
 deformation of the ADHM moduli space that  removes the small instanton
 singularity \refs{\nakajima}.
Such effects, which
 will not be discussed  very much in this paper, should generate interactions with fewer
  derivatives than those considered here.

\newsec{The low energy dynamics of D3-brane excitations}

At long wavelengths the dynamics of the excitations on a
D$p$-brane may be well approximated by an action that consists of
the sum of the Dirac--Born--Infeld (DBI) action and a Wess--Zumino
(WZ) term,
\eqn\dthreeact{S_p = S^{DBI}_p + S^{WZ}_p,}
The DBI term  takes the form
\eqn\binfa{S_p^{DBI}
= T_{(p)} \int d^{p+1}x \;
\sqrt{\det\big((G_{\mu\nu}+B_{\mu\nu})\partial_m Y^\mu\partial_n
Y^\nu
 +2\pi \alpha'\,F_{mn}\big)},}
where $F_{mn}$ is the world-volume Born--Infeld field strength,
 $G_{\mu\nu}$  is the target-space metric,  the tension $T_{(p)}$ is given by
\eqn\tenp{T_{(p)}= 2\pi(4\pi^2\alpha')^{-(1+p)/2}\,   e^{-\phi} ,}
 the antisymmetric tensor potential,  $B_{\mu\nu}$, will be set
 equal to zero in the remainder of this paper.
With this definition the   Yang-Mills   coupling constant has the
value
\eqn\gpdef{g^2_{p+1} =4\pi (4\pi^2\alpha')^{(p-3)/2}\, e^\phi,}
as can be seen from the expansion of \binfa\ to order $F^2$.
  Our conventions are that ten-dimensional space-time
vectors  are labeled by  $\mu =0,\dots,9$ while a $SO(9,1)$
Majorana--Weyl spinor will be labeled  $\calA = 1,\dots 16$. The
world-volume directions are $m, n ,\dots = \mu = 0,\dots, p$,
while the   directions transverse to the D$p$-brane
 will be labeled  $a, b, \dots = p+1,\cdots,9$.
 The low energy limit of the
 WZ term in the action \dthreeact\ can be deduced
by requiring the absence of chiral anomalies in an arbitrary
configuration of intersecting D-branes
\refs{\bainbachasgreen,\douglasa,\greenharveymoore,\ooguristudents}.
In static gauge the tangential components of the embedding
coordinates are identified with the world-volume coordinates,
while the normal components are identified with the scalar
world-volume fields.
\eqn\staticg{Y^m(x^m)= x^m,\qquad Y^a(x^m)= 2\pi \alpha' \;\varphi^a(x^m)}
Where $\varphi^a$ is a canonically normalized scalar field of dimension
$[L]^{-1}$, (whereas the coordinate $Y$ has dimension $[L]$). More
generally, one defines normal and tangent frames as summarized in
appendix B.

The DBI  action is exact only in the approximation that all
derivatives of the field strength $\partial F$ and second
derivative (acceleration) terms on scalars $\partial^2 \varphi$
are small enough to be ignored \refs{\fradkintseytlin}.  Terms of higher
order in derivatives may  be studied by considering the scattering
of open strings   (or ripples) propagating on the D-brane. The
tree-level amplitude for the scattering  of four ground-state open
superstrings
 on a D$p$-brane with any permitted value of $p$ is given by
compactification of the
  familiar expression,
\eqn\treeone{A^{tree}_4= {\cal N}
K\, \Big\{ {\Gamma(-\alpha' s)\Gamma(-\alpha' t)\over
    \Gamma(1-\alpha' s- \alpha' t)}+(s \leftrightarrow u)+ (t
\leftrightarrow u)\Big\}, } where the kinematic factor $K$ is of
the form
\eqn\kininv{K= -16
t_8^{\mu_1\nu_1\mu_2\nu_2\mu_3\nu_3\mu_4\nu_4}\zeta^{(1)}_{\mu_1}k^{(1)}_{\nu_1
 }
\zeta^{(2)}_{\mu_2}k^{(2)}_{\nu_2}\zeta^{(3)}_{\mu_3}k^{(3)}_{\nu_3}
\zeta^{(4)}_{\mu_4}k^{(4)}_{\nu_4},} where $\mu,\nu =0,\cdots,9$
and $t_8$ is a well-known eighth-rank tensor that is an
appropriately symmetrized sum of products of four Kronecker
$\delta$'s \refs{\greenschwarza}. The normalization factor is given by,
\eqn\normdef{{\cal N}   = -{1\over 16\pi} {\alpha'}^2e^{-2\phi}.}

In order to  apply this formula to the
scattering of open-string excitations of a
 three-brane in static gauge,
 one simply restricts the momenta $k^{(r)}_{\mu_r}$ to lie in the
 world-volume directions for which $\mu_r=a_r= 0,1,2,3$. The
 polarization vectors
  of the world-volume vector field are denoted
 $\zeta^{(r)}_{a_r},a_r=0,1,2,3$
   whereas the six scalars $\varphi$ have wave functions   $\zeta^{(r)}_a$, $a=
\mu-3 = 1,\cdots,6$.
For the transverse scalars the kinematic factor is particularly simple
\eqn\kinfac{K=-16\Big(  st \zeta_1 \cdot\zeta_3\;
\zeta_2\cdot\zeta_4+
tu \zeta_1\cdot\zeta_2\; \zeta_3\cdot\zeta_4+su \zeta_2\cdot\zeta_3\;
\zeta_1\cdot\zeta_4 \Big).}

Using the expansion of the logarithm of the Gamma function
\eqn\gamexp{\ln\Gamma(1+z)= -Cz+ \sum_{k=2}^\infty (-1)^k {z^k\over
    k}\zeta(k),}
and using $s+t+u=0$ it is easy to see that the  ratio of $\Gamma$'s in
\treeone\ has the expansion,
\eqn\gammatwo{\eqalign{ {\Gamma(-\alpha' s)\Gamma(-\alpha' t)\over
    \Gamma(1-\alpha' s- \alpha' t)}
&= {4\over {\alpha'}^2 st} \exp\Big( -{\zeta(2)\over 4} {\alpha'}^2 st -
    {\zeta(4)\over 32}{\alpha'}^4 st(2t^2+2s^2+3st)+o(s^4)\Big)\cr
&= {4\over  {\alpha'}^2 st} -\zeta(2) + {\zeta(2)^2\over 8}  {\alpha'}^2
st-
 {\zeta(4)\over
    8} {\alpha'}^2 (2t^2+2s^2+3st)+o(s^4)}.}
The first term on the right-hand side describes
massless tree level exchange that arises
    from the Yang--Mills action that is obtained as the leading term in
    the expansion of $S_3$ in powers $F$.   However, these  pole terms
cancel in
    the abelian case  of relevance to us  after the three terms in
\treeone\ are added,
     since the scattering states carry no charge. The
second term
    on the right-hand side of \gammatwo\
    corresponds to the  $F^4$ term in the expansion of the
    DBI action while  the third and fourth terms
    correspond to higher derivative interactions of the form $(\partial
F)^4$ and $(\partial^2 \varphi)^4$. The four-point amplitude has a series
expansion
in powers of  spatial derivatives of the form
\eqn\asum{A^{tree}_4
= A_4^{(0)} + {\alpha'}^2 A_4^{(2)} + {\alpha'}^4 A_4^{(4)} + \dots .}
After adding the three different orderings,  the $A_4^{(4)}$ terms in
the  amplitude for scattering  four transverse scalars are given by
\eqn\ordertwo{\eqalign{A_4^{(4)}&= {{\cal N} K\over {\alpha'}^2}
\,  \left( {\zeta(2)^2\over 8} + {5\zeta(4)\over
    8}\right)(st+tu+su)\cr
& = -  {\pi^3e^{-\phi}\over 3\times 2^{6}  }\,
(s^2+t^2+u^2)\,\big(st \zeta_1 \cdot\zeta_3\;
\zeta_2\cdot\zeta_4+
tu \zeta_1\cdot\zeta_2\; \zeta_3\cdot\zeta_4+su \zeta_2\cdot\zeta_3\;
\zeta_1\cdot\zeta_4 \big).}}

\subsec{Higher derivative terms and $SL(2,Z)$ invariance}

A consequence  of  S-duality of IIB superstring theory  is that the
 D3-brane is inert under $SL(2,Z)$ transformations which
act on $\tau$ by
\eqn\sltwoact{\tau \to {a\tau + b \over c\tau +d},}
where $a,b,c,d \in Z$ and $ad-bc=0$.  The metric tensor
 in the Einstein frame is inert under this transformation while the
 antisymmetric two-form potentials transform as a $SL(2,Z)$
 doublet.

Such an S-duality transformation is a symmetry of  the  equations of
motion
for the D3-brane that come from the variation of the
sum of the D3-brane
DBI  and  WZ actions,
when it is   accompanied by a
$SL(2,Z)$ electromagnetic duality transformation of the world-volume
fields \refs{\tseytlina,\greenguta}.  The field strength, $F$, together
 with its dual, $G^{mn}= i \delta S_3 / \delta F_{mn}$, form an
 $SL(2,Z)$ doublet  which  transforms as,
\eqn\emdual{ \pmatrix{*G\cr F}\to
  \pmatrix{a&b\cr c&d}\pmatrix{*G\cr F}.}
 The  combinations
\eqn\aself{F^+ = {1\over i \tau_2} (\tau F - *G),\qquad F^- = {1\over i
\tau_2}
(\bar\tau F - *G)}
transform  as  forms of weight $(1,0)$ (for $F^-$) and $(0,1)$ (for
$F^+$)\foot{In this notation   a
modular form of weight $(p,q)$ transforms with holomorphic weight $p$ and
anti-holomorphic weight $q$.} so that
\eqn\aselftrans{F^- \to (c\tau + d)\, F^-,\qquad F^+ \to
(c\bar \tau + d)\, F^+, }
which  means that
the combination
\eqn\invcom{\tau_2 F^+ F^-}
is invariant under $SL(2,Z)$.
 At lowest order in the low energy expansion $G^{mn}$ is given by its
 Maxwell form, $G_{mn} =
ie^{-\phi}  F_{mn} + C^{(0)} {1\over 2} \epsilon_{mnpq}F^{pq}$  and the
expressions $F^\pm$ are the
self-dual and anti self-dual field strengths,
\eqn\dualdef{F^\pm = F \pm * F.}
These formulae are appropriate for euclidean  signature for which
$**=1$.

The preceding discussion of modular invariance only applies in the
approximation that the Born--Infeld field $F$ is constant so that the
action $S_3$ is valid.   However,
the issue of $SL(2,Z)$ invariance of the D3-brane when $F$
is not constant, so that the derivative of $F$ is non-zero,
 requires further investigation.
The low-energy expansion of the four-point function
\ordertwo\  determines such   higher derivative
corrections.  The first terms
(with  bosonic open-string fields) that arise beyond the $F^4$ terms have the
schematic form,
\eqn\higherder{S'_3=
{ \pi^3{\alpha'}^4\over12}\int d^4x \sqrt{g}\, \tau_2 \Big( (\partial^2
  \varphi)^4 +  (\partial^2
  \varphi)^2\, \partial F^+\partial F^- + (\partial F^+)^2
(\partial F^-)^2  \Big),}
where the exact tensor structure of  contractions
 of each  term is  determined by demanding that \higherder\ reproduces
 \ordertwo\ when transformed into momentum space. The canonically
  normalized  scalar field $\varphi$  is related to the embedding
coordinate  $Y$ by \staticg.

In order to exhibit the transformation properties of these higher
derivative terms  under
$SL(2,Z)$ it is convenient  to transform the world-volume metric  to
the Einstein frame using
$g_{\mu\nu} = \tau_2^{-1/2}g^{(E)}_{\mu\nu}$, where $g^{(E)}_{\mu\nu}$
is the  modular invariant Einstein frame metric.
In the Einstein
frame  \higherder\ can be expressed as
\eqn\highereinst{S'_3= { \pi^3{\alpha'}^4\over12}\int d^4x \sqrt{g^{(E)}}\,
\tau_2 \Big(
(\partial^2
  \varphi)^4+ \tau_2(\partial^2\varphi)^2 \, \partial F^+\partial F^-
+\tau_2^2 (\partial F^+)^2
(\partial F^-)^2   \Big).}
 In order to avoid complications we will
 specialize to the case in which $\tau$ is constant.
The equations of motion that follow  from the total action $S_{BI}+S_{WZ}+S'_3$ fail to be
$SL(2,Z)$ invariant only because of the overall factor  $\tau_2$  in
$S'_3$.   To see this it is important to note that
the higher derivative terms in \highereinst\ modify the expression for
$G$ and hence the transformation rules \emdual\ to  the
fixed order of $\alpha'$ we are considering. This means that in order for
the full
nonperturbative
expression to be modular invariant this overall  factor of $\tau_2$
must be replaced
by a modular function, $h(\tau,\bar\tau)$, leading to
\eqn\fourfmod{S'_3= {\pi^2{\alpha'}^4\over 4}\int d^4 x \sqrt{g^{(E)}}
h(\tau,\bar\tau)
 \Big
( (\partial^2\varphi)^4+ \tau_2 (\partial^2\varphi)^2 \,\partial
F^+\partial F^-
+\tau_2^2  (\partial F^+)^2
(\partial F^-)^2 \Big).}

Apart from the tree-level contribution to these four-string processes
there is also
a logarithmically divergent  one-loop contribution.  This
arises, for example, in a field theory calculation of  the  $\langle
F^2(x_1) F^2(x_2)\rangle$ correlation
function, in which the
one-loop diagram has two $F^4$ vertices \refs{\shmakova,\zanon}.  Similarly,
it is straightforward to see that
there is a logarithmic
 infrared divergence in the open string one-loop amplitude.
 These observations make it  plausible that the
function $h$ is proportional  to
\modfun\  which  has the
the weak coupling   expansion \expanh.
 Support
for this ansatz is reinforced by the fact that
the interactions in \fourfmod\ combine
naturally with the induced $($curvature$)^2$ terms discussed in
 \refs{\bainbachasgreen} for which the prefactor $h$ was  proportional
to \modfun.    The presence of terms in \fourfmod\ of the form $(\partial^2
\varphi)^4$
in the D3-brane action  have a natural geometric
origin in terms of the contribution of the second fundamental form to
the action for a D3-brane embedded in a curved target space, as will be
seen in section 7.

\newsec{The D3-brane in the presence of a D-instanton}

\subsec{Collective coordinates}

The conventional instanton moduli space  arises in the Higgs
branch of the gauge theory  and  has a natural D-brane interpretation
\refs{\wittenmoduli\douglasa}.
In the following we will be describing the case of a single D3-brane
so the   world-volume  gauge
theory is abelian and  there is no Higgs branch and no `fat'
instantons.
However, within string theory there is a well-defined prescription for
describing  the moduli space of such a
 `pointlike'  instanton in terms of the configuration of
open strings joining the D3-brane to a D-instanton at a fixed transverse
separation
\refs{\greentwob,\polchdinst,\gutbound,\mli}.

Following \refs{\doreythree,\wittenmoduli,\douglasa} we may consider the
D-instanton/D3-brane system
to be obtained by T-duality from the D5/D9 system compactified on a
six-torus.
The $SO(10)$ of the euclidean ten-dimensional theory is broken in
this background to  $SO(6)\times SO(4)\sim
SU(4)\times SU(2)_L\times SU(2)_R$,
 where the
$SO(6)\sim SU(4)$ is the rotation group in the D5-brane which is the
R-symmetry group
of $\calN=4$ Yang--Mills theory, while the
 $SO(4)\sim SU(2)_L\times SU(2)_R $ is
the rotation group in the directions transverse to the
D5-brane.\foot{The defining representations of
$SU(4)\times SU(2)_L\times SU(2)_R$ will
be labeled by the indices $A=1,2,3,4$, $\alpha, \dot \alpha=1,2$.}
Making   six T-dualities in the directions $\mu=4,\cdots,9$ transforms
the D9-brane
into a D3-brane  filling  the $\mu= n = 0,1,2,3$
directions. After turning on
Wilson lines the
D-instanton is located at a space-time point separated from the D3-brane
in the directions  $a=  \mu-3 = 1,\dots,6$.  We
will label the ADHM supermoduli using standard  notation (as, for
example, in \refs{\doreythree,\doreytwo}).

The dynamics of a single  D$p$-brane  is determined by
 the reduction to $p+1$ space-time dimensions  of the ten-dimensional
 supermultiplet of
 Maxwell theory, $(A_\mu,\Psi^{\cal A})$, where ${\cal A}$ denotes the
 sixteen components
 of
 a Majorana--Weyl spinor.\foot{In
the following the euclidean continuation of a
 ten-dimensional Majorana fermion is chosen to be  real.    Since
 the conjugate spinor is  not real, this
 leads to a non-hermitian euclidean hamiltonian, which is not a problem
  since hermiticity is not relevant in euclidean
 fields theory.} Thus,
the massless world-volume fields of the D3-brane and the collective
coordinates of the D-instanton are given by  dimensional reduction of the
ten-dimensional  theory to four and zero dimensions respectively.
In  the case of the D3-brane the bosonic  fields comprise the
 six scalars and a gauge field
\eqn\dthreebos{A_\mu=({\varphi_{AB},A_n}).}
Here  $A_n$ transforms as
$({\bf 1,2,2})$ of $SU(4)\times SU(2)_L\times SU(2)_R$
while  $\varphi_{AB}$ transforms as $({ \bf 6,1,1})$,
where the six transverse scalars $\varphi^a,a=1,\cdots,6$ are
related to
$\varphi_{AB}$ by
\eqn\phisuso{ \varphi_{AB}= {1\over \sqrt 8}  \Sigma^a_{AB} \varphi^a =
{1\over 2}
\epsilon_{ABCD} \bar \varphi^{CD}.}
(where the Clebsch--Gordon coefficients $\Sigma^a_{AB}$ are defined in appendix
A). The massless fermionic D3-brane fields are given by the spinors
\eqn\dthreeferm{\Psi=(\Lambda^A_{\alpha},\bar\Lambda_A^{\dot\alpha}),}
where the gauginos $\Lambda^A_\alpha,{\bar\Lambda}_A^{\dot\alpha}$,
transform as $\bf{(4,2,1)}$ and $\bf{(\bar{4},1,2)}$, respectively.

Similarly  the collective coordinates of an isolated D-instanton
are determined by the reduction of super Yang--Mills
to zero space-time dimensions.
The bosonic vector potential  decomposes under $SO(4)\times SO(6)$ as
\eqn\adecom{A_\mu = (a_n, \chi_a),}
($n=0,1,2,3$ and $a=1,\cdots,6$).  In terms of the covering group
the bosonic fields are
$\chi^a$, which  forms a  ${\bf (6,1,1)}$, and
 $a_{\alpha\dot \beta} = \sigma^n_{\alpha\dot \beta}
 a_n$ which forms a ${\bf (1,2,2)}$.
  The vector $\chi^a$ is  also conveniently written as a $4\times 4$
antisymmetric
matrix, $\chi_{AB}$, which is defined by
\eqn\chiant{\chi_{AB} = {1\over \sqrt 8} \Sigma^a_{AB} \chi_a.}
The D-instanton  fermions  arise from the ten-dimensional Majorana--Weyl
 spinor which decomposes as  ${\bf 16}
\to (
{\bf 4,2,1}) \oplus ({\bf \bar 4,2,1})$ so that
\eqn\fermdec{\Psi = (M^{\prime A}_\alpha, \lambda_{A\dalp}).}

Although the action for a single D5-brane is trivial it becomes non-trivial
in
the presence of the D9-brane, which breaks the six-dimensional $(1,1)$
supersymmetry
to $(0,1)$.  The interactions arise due to the presence
of  massless fields associated with the open  strings joining
the D5-brane and the D9-brane.  These are the $(0,1)$
hypermultiplets $(\mu^A,w_\dal)$ and
$(\bar \mu^A,\bar w^\dal)$  which consist of the
bosonic fields $w_\dal$ and $\bar w^\dal = (w_\dal)^*$  in
$({\bf 1,1,2})$ and their fermionic partners $\mu^A$ and $\bar\mu^A$
 in $({\bf 4,1,1})$.
After  T-duality on the six-torus the separation of the D3-brane at the
origin
 and the D$(-1)$-brane is given by $\chi^2 \equiv \chi^a\chi_a = L^2$,
which is the length of the
 stretched strings joining the two D-branes.

In addition to these fields it is usual to introduce the
anti-self-dual
auxiliary  field, $D_{mn} = -(*D_{mn})$,
in order to make supersymmetry
manifest.
The self-duality property means that this field transforms in the adjoint
of $SU(2)_R$ and can be rewritten in terms of $D^c$ ($c=1,2,3$) using
$D_{mn} = - D^c \bar\eta_{mn}^c$,   where $\eta^c_{mn}$ is the 'tHooft
symbol defined in appendix A.

The form  of the action for the D-instanton collective coordinates
follows from the toroidal compactification of the $d=6$ $N=1$ world-volume
theory of a single D5-brane in the presence of a D9-brane and has the form
 (see, for example, \refs{\doreythree} with  $k=N=1$),
\eqn\dtermb{S_{-1}= -2\pi i \tau + {1\over g_0^2} D_c^2 + iD^c W^c
 +\chi^2 W_0 - 2i
\bmu^A  \mu^B \chi_{AB}
 + i \left(\bmu^Aw_\dal \lambda_A^\dal
 + \bar w_\dal \mu^A \lambda_A^\dal   \right),}
where
\eqn\wdefs{W_0= \bar w w, \qquad W^c =  \bar w \bar\sigma^{mn } w\,
\bar \eta^c_{ mn}
\equiv  W^{mn}\,  \bar \eta^c_{ mn},}
(where $W^{mn} =  \bar w
\bar\sigma^{mn } w$)
which satisfy the constraint
\eqn\consw{W_0^2 = (W^c)^2 = W_{mn}W^{mn}.}
The quantity   $g_0$ in \dtermb\ is the zero-dimensional coupling
constant, which is given from \gpdef\ by
\eqn\relcoup{g_0^2 = 4\pi (4\pi^2 \alpha')^{-2}\, e^\phi ,}
and $\tau$ is the constant value of the complex bulk scalar field.
 It is notable that there are no couplings of the $a_n$ and
${M'}^A_\alpha$ in the action \dtermb, which arises from the fact that
these are the supermoduli associated with the relative longitudinal
position of the D-instanton and the D3-brane.

The auxiliary field $D^c$ may be integrated out,  producing
a factor  of $g_0^3$ in the measure which will be important
later. The  resulting moduli-space action  of the abelian D-instanton
becomes
\eqn\dterm{S_{-1}  = -2\pi i \tau +   {g_0^2 \over 4} (W^c)^2+
  \chi^2  W_0 - 2i
\bmu^A \mu^B \chi_{AB}
 + i \left(\bmu^A\, w  \lambda_A
 +\mu^A\,  \bar w  \lambda_A   \right).}
 In order to
recover the usual limit of  $\calN=4$ superconformal
 Yang--Mills theory in four dimensions it is
necessary to decouple the closed string sector \refs{\malda}  by taking the low
energy limit
$\alpha'\to 0$. Since $g_0^2 \sim e^{\phi} /\alpha'$
this  means taking  $g_0^2 \to \infty$ with fixed string coupling $e^\phi$
which enforces the condition
$w_\dal=\bar w_\dal=0$.  Therefore
the instanton induced terms to be discussed later vanish in this limit.
  This is
in accord with the fact that the instanton induced terms have higher
derivatives and vanish for dimensional reasons
in the $\alpha'\to 0$ limit.
We are here interested
in a more general situation in which the gravitational sector does not
decouple.

\subsec{Broken and unbroken supersymmetries}

The combined D-instanton/D3-brane system  has bosonic zero modes
associated with
 broken translation symmetries.  These comprise the ten modes
corresponding to the
 overall  space-time
translations of the composite system together with  four zero modes
that come from the invariance of the system under shifts of the
D-instanton alone  in the directions parallel to the D3-brane.
  Correspondingly  a fraction of the
thirty-two supersymmetry components are broken in this background.
Sixteen of these
broken supersymmetry components are superpartners of the  broken
overall translational
symmetry.  A further eight broken supersymmetry components are
superpartners of
the remaining  four broken translations.

These statements are encoded in the supersymmetry algebra starting from the
ten-dimensional Maxwell supersymmetry transformations,
given by
\eqn\supalg{\eqalign{\delta A_\mu & = i\bar \eta
\Gamma_{\mu}\Psi, \cr \delta \Psi & = -
\Gamma^{\mu\nu} F_{\mu\nu}\eta + \epsilon,\cr}}
where $\eta$ and
$\epsilon$ are two sixteen-component Majorana--Weyl spinors.  After
reduction to $p+1$ dimensions this
algebra describes the supersymmetries of a D$p$-brane background.  The
parameter $\eta$ labels the sixteen unbroken supersymmetries while
$\epsilon$ corresponds to the sixteen zero modes  generated by the
action of the broken supercharges on the background.  The latter are
the superpartners of the translational zero modes.

In the absence of the D-instanton the D3-brane is invariant under the
action
of ${\cal N}=4$ supersymmetry, which has $SU(4)$ as its R-symmetry group.
 Although the fully supersymmetric
generalization
of the Born--Infeld action is known \refs{\susybia,\susybib,\susybic} we
need only
consider the transformation of the low energy Maxwell system  here.   The
action of the supersymmetries on the fields is
 given by the familiar $\calN=4$ algebra,
\eqn\transone {\eqalign{
\delta \bar\varphi^{AB} &= {1\over 2} (\Lambda^{\alpha\, A}
    \eta_{\alpha}^{B}-
        \Lambda^{\alpha\,B} \eta_{\alpha}^{A}) + {1\over 2}
        \varepsilon^{ABCD}
        {\overline \xi}_{\dot \alpha\, C}
        {\overline \Lambda}^{\dot \alpha}_{D} \cr
        \delta \Lambda_{\alpha}^{A} &= - {1\over 2}  F^-_{m n}
        {{\sigma^{m n}}_{\alpha}}^{\beta} \eta_{\beta}^{A} + 4i
        \, {\partial}_{\alpha {\dot \alpha}} \bar\varphi^{AB}\,
        {\overline \xi}^{\dot \alpha}_B \cr
        \delta A_m &= -i \Lambda^{\alpha \, A}
        {\sigma^m}_{\alpha \dot \alpha}
        {\overline \xi}^{\dot \alpha}_A
        -i  \eta^{\alpha \,A} {\sigma^m}_{\alpha \dot \alpha}
         {\overline \Lambda}^{\dot \alpha}_{A}\cr
          \delta \bar\Lambda^{\dot \alpha}_A &= - {1\over 2}  F^+_{m n}
        {{\bar\sigma^{m n}}_{\ \ \ \dot \beta}}\ ^{\dot\alpha} \bar
        \xi^{\dot \beta}_A  - 4i
        \, {\partial}^{\dot \alpha  \alpha} \varphi_{AB}\,
        \eta_{ \alpha}^B,
 \cr }}
where lower indices $A, B =1,2,3,4$ label the defining representation of
$SU(4)$  or, equivalently,  a chiral spinor representation
of $SO(6)$ (while upper indices denote the conjugate representations).
The  Grassmann variables  $\bar\xi_A^{\dot\alpha}$ and
$\eta^A_\alpha$ parameterize the sixteen unbroken supersymmetries
that descend from $\eta$ in \supalg\ and
are chiral spinors of both $SO(6)$ and $SO(4)$,
\eqn\grassmp{{1\over \sqrt{2\pi}} \pmatrix{1\cr 0\cr} \otimes \pmatrix{0
\cr
\bar \xi_A^{\dot \alpha}\cr}, \qquad
 {1\over \sqrt{2\pi}} \pmatrix{0\cr 1\cr} \otimes \pmatrix{\eta^A_\alpha
\cr
0\cr},}
where $SO(4)$ chirality is denoted, as usual, by dotted and undotted
indices.

Likewise, in the absence of the D3-brane, the D-instanton is invariant
under the sixteen supersymmetry transformations,
\eqn\transtwo{\eqalign{
\delta \bar \chi^{AB} & ={1\over 2} ({M'}^{\alpha\,A}
    \xi_{\alpha}^{B}-
        {M'}^{\alpha \, B} \xi_{\alpha}^{A}) +   {1\over 2}
        \varepsilon^{ABCD}
        {\overline{\xi}_{\dot \alpha}\, C}
        {\lambda}^{\dot \alpha}_{D} \cr
          \delta a_m  &= -i
         {M'}^{\alpha \, A}
        {\sigma^m}_{\alpha\dot \alpha}
        {\overline \xi}^{\dot \alpha}_{A}
        -i  \xi^{\alpha \,A} {\sigma^m}_{\alpha \dot \alpha}
        {  \lambda}^{\dot \alpha}_{A}, \cr}}
where $\bar \xi_A^{\dot \alpha}$ and $\xi^A_\alpha$ are the sixteen
        supersymmetry parameters.
The notation has been chosen to emphasize the fact that in the coupled system
of a D3-brane in an instanton background eight of the supersymmetries
in \transone\ and \transtwo\ are common and are therefore unbroken in
the composite system.  These are the ones associated with the
parameters $\bar\xi_A^{\dot\alpha}$.  The unbroken $(0,1)$
supersymmetry acts on the
$w$ and $\mu$   by the transformations
\eqn\transthree{\eqalign{
\delta w_{\dot\alpha}  & = \bar\xi_{A \dot\alpha} \mu^A \cr
\delta \mu^A & = - 4i w_{\dot \alpha} \bar\chi^{AB}\,
\bar\xi_B^{\dot \alpha} \cr
\delta \lambda^{\dot \alpha}_A & = {i\over 2}g_0^2\, \bar w w \,\bar\xi^{\dot
\alpha}_A ,
\cr}}
in addition to the $\bar \xi$ transformations in \transone\ and \transtwo.
As will be discussed later, in the presence  of D3-brane source
fields  there are other $\bar \xi$ transformations that modify the
transformations
in \transone.

The other twenty-four supersymmetry components are broken on one or
other of the D-branes or on both.
The quantity $M'$ is the goldstino for the supersymmetry that is
broken on the D-instanton and is associated
with the shift transformation,
\eqn\brokinst{\delta_\eta {M'}^A_\alpha = \eta^A_\alpha,}
with the remaining D-instanton coordinates and twist fields being inert.
The field $\Lambda$ is the goldstino for the supersymmetry that is
broken on the D3-brane but not on the
D-instanton which is implemented by the shift,
\eqn\brokthree{\delta_\xi {\Lambda}^A_\alpha = \xi^A_\alpha,}
leaving  $w$ and $\mu$ and the other D3-brane fields inert.
The final set of broken supersymmetries are broken on  both the
D3-brane and on the D-instanton and are associated with the
transformations,
\eqn\brokboth{\delta_\rho \lambda_A^{\dot \alpha} = \rho_A^{\dot \alpha} =
\delta \bar \Lambda_A^{\dot \alpha},}
where $\rho_A^{\dot \alpha}$ are eight further Grassmann variables.
None of the other fields transform under these  supersymmetry
components.

The following table summarizes the various broken and unbroken
supersymmetries (denoted $b$ and $u$, respectively),
\eqn\matsu{\matrix{ &\bar \xi_{\dot\alpha A}&\xi_\alpha^A&\rho_{\dot\alpha
 A}&\eta_\alpha^A\cr
D3& u&b& b&u\cr
D(-1)&u &u &b& b} .}
 The twenty-four Grassmann parameters for the broken
supersymmetries are the supermoduli of the system.

\subsec{Vertex Operators and disk diagrams}

The interactions between the open strings summarized by \dterm\ can be
obtained by
considering the insertion of the various vertex operators on a disk
which has a segment of
purely Dirichlet boundary (the D segment) and a sector with Neumann
conditions in the $m=1,2,3,4$ directions (the N segment). These
boundary conditions require the presence of two twist operators that
reside at the points at which the boundary conditions flip.
This will later be seen to be a very useful way
of summarizing
the interactions of external D3-brane open-string sources.

The vertex operators attached to a Dirichlet boundary are those for the
moduli of an isolated D-instanton and represent the interactions of
open strings
with endpoints satisfying  purely Dirichlet (DD)
boundary conditions.\foot{From now on we
will mainly use
language of the D-instanton/D3-brane system although we will also make
reference to the related D5-brane/D9-brane system.}
They are given  (in the $-1$ picture) by
\eqn\vertone{V_{-1}(\chi^a)= \chi_a\; e^{-\phi}\psi^a,\qquad
V_{-1}(a_n)=
  a_n \;e^{-\phi}\psi^n, }
where here and in the following the subscript of the vertex
operator denotes its superghost number.
The vertex operators for the fermionic fields are (in the
$-1/2$  picture)
\eqn\verttwo{V_{-1/2}(\lambda)= \lambda^{\dot{\alpha}}_A\;
  e^{-\phi/2}\Sigma_{\dot{\alpha}}\Sigma^A,\qquad V_{-1/2}(M^\prime)=
  M^{\prime\;A}_\alpha \;e^{-\phi/2}{\Sigma}^\alpha \Sigma_A,}
where  a  $SO(10)$ spin field $\Sigma^a$
$a=1,\cdots,16$ is decomposed into a
product of a $SO(4)$ spin fields $\Sigma_a, \Sigma_{\dot{a}}$ and a
$SO(6)=SU(4)$ spin fields $\Sigma_A,\Sigma^A$.

There are no interactions involving the
strings that begin and end on the D-instanton.
The terms in
\dterm\ arise from the disk with a portion of the boundary having
Neumann conditions in the transverse directions
and a portion having Dirichlet conditions.  With these conditions two
of the vertex
operators must be twist operators.
In the presence of a D3-brane there are additional collective coordinates
which come from the strings stretched between the D3-brane and the
D-instanton which
have Neumann conditions in the $m=0,1,2,3$ directions at one end  and
Dirichlet in all
other directions (DN strings)
These  descend  from the  hypermultiplets
in six dimensions.
The vertex operators for $w_{\dot\alpha}$ and $\mu^A$ are given (in the
ghost number $-1$ picture) by
\eqn\vertthreea{V_{-1}(w)= w_{\dot{a}} e^{-\phi} \Delta
\Sigma^{\dot{a}},\qquad
V_{-1/2}(\mu)= \mu^A e^{-\phi/2}\Delta \Sigma_A,}
where $\Delta=\sigma_0\sigma_1\sigma_2\sigma_4$ is the product of $Z_2$
twist fields  which twists the bosonic field $X^\mu$ so that it
interpolates between
 Neumann and Dirichlet boundary conditions in the $\mu=0,1,2,3$
directions.

The various terms in \dterm\ follow simply
by evaluating three-point  and four-point functions of vertex operators on
the disk,
 ensuring
that the total
superconformal ghost number is always $-2$.  For example,
\eqn\vlam{\langle c V(\lambda) c V(w) c V(\bmu)  \rangle =
 i\pi \,\bmu^Aw_\dal \lambda_A^\dal,}
where $c$ is the superconformal ghost.   The other terms in \dterm\
follow in a similar
manner.

In addition to fields localized at the instanton there are open-string
excitations on
the three-brane world-volume. The  massless bosonic excitations living on
the three-brane and carrying longitudinal momentum $k^m$
 are described by vertex operators  in the
$-1$ superghost  picture  of the form
\eqn\threeboss{V_{-1}(A)= A_n e^{-\phi} \psi^n e^{ik \cdot X},\qquad
V_{-1}(\varphi)=
\varphi_a
  e^{-\phi} \psi^a  e^{ik \cdot X} ,}
while the  gaugino  vertex operators in the $-1/2$ picture are given by
\eqn\verttwob{V_{-1/2}(\bar{\Lambda})= {\bar\Lambda}_A^{\dot\alpha}\;
  e^{-\phi/2}\Sigma_{\dot{\alpha}}\Sigma^A\,  e^{ik \cdot X},\qquad
V_{-1/2}(\Lambda)=
  \Lambda^A_\alpha \;e^{-\phi/2}{\Sigma}^\alpha \Sigma_A\,  e^{ik \cdot X}.}

\newsec{ Inclusion of D3-brane sources}

We would now like to generalize the moduli space action \dterm\ to
include sources that correspond to the couplings of the open-string
ground states of the D3-brane.  The resulting action should encode the
supersymmetry transformations \supalg-\brokboth\ and will lead to an
evaluation of the leading effects of the D-instanton on open-string
scattering on the D3-brane.
In order to include these sources we need to consider the insertion of
vertex operators  on the N segment of the boundary of the disk.  Such
an operator describes an on-shell plane-wave scattering state with
momentum $k^m$.

The simplest  diagram of this type  is given by the insertion
a three-brane  gauge field vertex and two bosonic twist fields
without any $M'$ insertions,
 \eqn\simpsource{\langle F^+\rangle_{w\bar w}
=\langle cV_0(A_m)cV_{-1}(\bar w)cV_{-1}(w)\rangle=
W^{mn}\,  F_{mn}^+ \equiv W^c \bar \eta_c^{mn}
 F^+_{mn},}
which is proportional to the self-dual part of the Maxwell field.
The wave function $F^+$ includes a plane-wave  factor $e^{ik\cdot a}$,
where the collective coordinate $a^m$ will later be integrated.
This is the basic one-point function from which other one-point
functions for massless  D3-brane states can be obtained by considering
 processes involving a single D3-brane vertex
operator $V_g(\Phi)$ (where $\Phi$ is any D3-brane open-string
ground state and $g$ is the superghost
number of the vertex)  inserted onto a disk with two bosonic twist
operators and with $n$ insertions of integrated $M'$ vertex
operators on the D segment of the boundary.  We will denote this
process by the symbol
\eqn\notexpe{\langle \Phi \rangle_{w\bar w;n} =
\langle  c V_g(\Phi) c V_{-1}(\bar w) c V_{-1}(w)  \int_D V_{l_1} (M') \dots
V_{l_n}(M')\rangle ,}
where $\int_D$ indicates that the integration is over the Dirichlet sector
of the
boundary and $l_r =\pm 1/2$ are the ghost numbers of the $M'$ vertex operators.
Since the two $V(w)$ vertices saturate the
ghost number anomaly for the disk the total ghost number for the other
vertex operators in \notexpe\ has to be zero, so that $g + \sum_{r=1}^n l_r
=0$.
The quantity \notexpe\  is easily evaluated in terms
of the expression \simpsource\ as   follows from the
transformation properties of the D3-brane vertices under the
$\eta$ supersymmetries,
\eqn\susytra{\eqalign{V_0(\delta_\eta A_m) =&  [\eta Q_{+1/2},
V_{-1/2}(\bar\Lambda)], \qquad
V_{-1/2}(\delta_\eta \bar\Lambda)= [\eta Q_{-1/2},
V_0(\partial\varphi)],\cr
\qquad V_0(\delta_\eta
\varphi)= &  [\eta Q_{+1/2}, V_{-1/2}(\Lambda)],\qquad
V_{-1/2}(\delta_\eta \Lambda)= [\eta Q_{-1/2},
V_0(F^-)]  ,\cr}}
where  the quantities $\delta_\eta \Phi$ are the $\eta$ supersymmetry
variations
of \transone.  The supersymmetry charge $Q$ can be
represented as $\eta Q_{\pm 1/2}= \int_D V_{\pm 1/2}(\eta)$, where $V_{\pm
1/2}$ is the
same expression as the $M'$ vertex operator.

This can be used, for example, to
evaluate the diagram with one $\bar\Lambda$ vertex, two bosonic twist
fields and one $M'$ vertex,
 \eqn\firstdb{\eqalign{\langle \bar\Lambda  \rangle_{w\bar w; 1}
& = \langle
cV(\delta_{M'} A_m)cV(\bar w)cV(w)\rangle\cr
&=\langle cV(\bar\Lambda)cV(\bar w)cV(w) \int_D V(M')\rangle\cr &=-i
W^{mn}\,  {M{'}}^A \sigma_{[m}\partial_{n]}\bar\Lambda_A.\cr  }}
Successive application of the supersymmetry transformations parameterized
by
$\eta$ gives
\eqn\etaapp{\eqalign{\langle  \varphi_{AB}\rangle_{w\bar w;2}
&=4 W^{mn}\;
{M'}^B \sigma_{mp}{M'}^A \partial_n \partial^p \varphi_{AB}\cr
\langle  \Lambda\rangle_{w\bar w; 3}&= 2  W^{mn}\;\epsilon_{ABCD}
{M'}^B
\sigma_{pm}{M'}^A {M'}^{C\, \alpha} \partial_{n} \partial^p
\Lambda^D_\alpha \cr
 \langle  A \rangle_{w\bar w; 4}&= W^{mn}\;\epsilon_{ABCD}
{M'}^B\sigma_{pm}{M'}^A\;
{M'}^C \sigma^{kl}{M'}^D\; \partial_n \partial^p F^-_{kl}.
\cr}}
The collection of terms \simpsource, \firstdb\ and \etaapp\ can be
combined into a superfield, $\Phi_{mn}(x^n,M') W^{mn}$
where,
\eqn\supone{\Phi_{mn}(x^n, M')  = F^+_{mn} +  i\Mp^{ A}
\sigma_{[m} \, \partial_{n]} \bar\Lambda_A
 + \dots .}
 The complete expansion for  this superfield is given in section
 4.1.  In a similar manner it is easy to construct the disk amplitudes
with two fermionic twist operators, $\mu$ and $\bar \mu$, which
combine into the expression $\Phi_{AB} \bar \mu^A\mu^B$ where,
\eqn\supthree{\Phi_{AB}(x^n, M') = \varphi_{AB} +{1\over 2}\epsilon_{ABCD}
 \Mp^{[C\alpha} \, \Lambda^{D]}_\alpha + \dots .}
 Finally, the disk  diagrams with one bosonic and one
 fermionic twist vertex combine into $\Phi_A^{\dot \alpha}(x^n, M')
w_{\dot\alpha} \mu^A $ where,
\eqn\suptwo{\Phi_A^{\dot \alpha}(x^n, M') =   \bar \Lambda_A^{\dot
\alpha} -4i \Mp_\beta^B\, \sigma^{n\, \beta\dot
\alpha}\, \partial_n\varphi_{AB} + \dots.}
The complete expressions for $\Phi_{AB}(x^n, M')$ and $\Phi_A^{\dot
\alpha}(x^n, M')$
are also given in section 4.1.   By construction,
these superfields are  invariant under the supersymmetry transformations
generated by shifts of $M'\to M' +\eta$,
 \eqn\susyone{\delta_{\eta} \Phi = \eta_\alpha^A {\partial \over \partial
  M^{\prime A}_\alpha } \Phi,}
provided  the component fields transform under the eight $\eta$
supersymmetry transformations of \transone.

The complete moduli space action, including the D3-brane sources,
can now be written as
\eqn\fullact{\eqalign{S_{-1}[\Phi_{mn}, \Phi_A^{\dot\alpha}, \Phi_{AB}] =
& -2\pi i \tau+ {1\over 4} g_0^2 (W^c)^2 + 2(\chi_{AB} -\Phi_{AB})^2 W_0 -i
\Phi_{mn} \bar \eta^c_{mn} W^c \cr
 & -2i (\chi_{AB}-\Phi_{AB})\bar \mu^A\mu^B -
 i(\lambda_A^{\dot \alpha} - \Phi_A^{\dot \alpha}) (\bar \mu^A w_{\dot
\alpha} + \mu^A \bar w_{\dot \alpha}),\cr}}
so that $S_{-1}[0,0,0] = S_{-1}$.
 The dimensions of all the fields in this
expression are determined by the fact that $g_0$ is proportional to
${\alpha'}^{-1}$.  In particular, this is consistent with the
assignment of the canonical dimensions to the free four-dimensional
bosonic and fermionic fields associated with the open strings of the
D3-brane which are
the leading terms in the expansion of the superfields $\Phi_{mn}$,
$\Phi_A^\dal$ and
$\Phi_{AB}$.

\subsec{Supersymmetries in the presence of D3-brane sources }

 The action \fullact\  is invariant under all thirty-two
supersymmetries provided the component D3-brane fields
transform in the appropriate manner.   These are the eight conserved
supersymmetries with parameter $\bar\xi^{\dot \alpha}_A$ and the
twenty-four supersymmetries with parameters
$\eta_\alpha^A$, $\rho_A^{\dot \alpha}$ and $\xi_\alpha^A$.
 In order to analyze this more completely it is useful to express the
 sources in terms of  $N=4$  on-shell superfields.

Recall \refs{\townsendsuper} that  the physical ${\cal N}=4$
 fields satisfying the linearized equations of motion are contained in a
superfield
 ${\cal W}_{AB}(\theta,\bar\theta)$, where the sixteen  Grassmann superspace
parameters are
 components of  $\theta^A_\alpha$ and
 $\bar\theta_A^{\dal}$ which transform  in the $({\bf 4,2,1})$ and
 $({\bf \bar 4,1,2})$ of $SU(4) \otimes SU(2)_L \otimes SU(2)_R$,
respectively.  Covariant derivatives $\bar D^A_\dal$ and $ D_A^\alpha$ are
 defined by
\eqn\covder{D^A_\dal = -{\partial\over \partial \bar \theta_A^\dal}
-i (\sigma\cdot
\partial)_{\dal \beta}  \theta^{A\beta}, \qquad
 \bar D_A^\alpha = {\partial\over \partial   \theta^A_\alpha} +i (\sigma\cdot
\partial)_{\alpha \dbet} \bar \theta^{A\dbet},}
while the supersymmetries are represented as
\eqn\susydefs{Q^A_\dal = -{\partial\over \partial \bar \theta_A^\dal}
+i (\sigma\cdot
\partial)_{\dal \beta}  \theta^{A\beta}, \qquad
\bar Q_A^\alpha = {\partial\over \partial   \theta^A_\alpha} -i
(\sigma\cdot
\partial)_{\alpha \dbet} \bar \theta^{A\dbet},}
The bi-fundamental superfield is defined to satisfy the constraints,
\eqn\conssup{\calW_{AB}= -\calW_{BA} = {1\over 2}
\epsilon_{ABCD}\bar \calW^{CD},
\qquad
D^C_\dal \calW_{AB} = \delta^C_{[A} \calW_{B]\dal}.}
It follows that the first few terms in the expansion of this superfield
have the form
\eqn\genexp{\eqalign{\calW_{AB}
& =   \varphi_{AB} + \bar\theta_{[A\dal} \, \bar\Lambda_{B]}^\dal
+{1\over 2} \epsilon_{ABCD}
 \theta^{\alpha[C} \, \Lambda^{D]}_\alpha -{1\over 4} \epsilon_{ABCD}
\theta^C\sigma^{mn} \theta^D \, F^{-}_{mn}\cr
&   -{1\over 2}
\bar \theta_A\bar \sigma^{mn}\bar \theta_B\, F^{+}_{mn}
 +i \theta^C \sigma^m\bar \theta_C\,              \partial_m
\varphi_{AB}
+ \dots.\cr}}
The derivative   superfield,    $\calW_B^\dal$, is defined in
 \conssup\ and can be written as a covariant derivative on $\calW_{AB}$,
\eqn\suplam{   \calW_B^\dal  ={3\over 2}   D^{A \dal} \calW_{AB} .}
Similarly, a tensor superfield  ${\cal W}_{mn}$ is obtained by applying a
further covariant derivative,
\eqn\supfs{\calW_{mn} =  D^A \bar \sigma_{mn}  D^B\, \calW_{AB}.}

The D3-brane fields that enter in the action $S_{-1}$ are functions of
$\Mp^A_\alpha$
that are identified as follows,
\eqn\sourcess{\Phi_{mn} = \calW_{mn}|_{\bar \theta =0, \theta = M'}, \qquad
\Phi_A^\dal = \calW_A^\dal |_{\bar \theta =0, \theta = M'}, \qquad
\Phi_{AB} = \calW_{AB}|_{\bar \theta =0, \theta = M'}.}
The fact that these are the correct identifications follows from the
fact that they  transform appropriately under the $\eta$ supersymmetry
transformations, which are given by $\eta^A_{\alpha}
 \bar Q_A^\alpha \, \Phi= -\eta^A_\alpha  (\partial/\partial
{M'}^A_\alpha)\, \Phi$.

The $\bar\xi$ supersymmetry transformation of $\Phi_{AB}$  is given
by
\eqn\susyxi{\eqalign{
\delta_{\bar\xi}  \Phi_{AB}   = \bar\xi_C\,
Q^C\, \calW_{AB}|_{\bar \theta =0, \theta = M'}
& =\bar\xi_C\,  D^C\, \calW_{AB}|_{\bar \theta =0, \theta = M'}
 - 2\bar\xi_C \sigma\cdot
\partial  \theta^C\, \Phi_{AB}\cr
& =  \bar\xi_{[A} \Phi_{B]} -   2\bar\xi_C \sigma\cdot
\partial  \theta^C \, \Phi_{AB} ,  \cr}}
while the $\bar\xi$ supersymmetry transformation
of $\Phi_A^\dal$  is  given by
\eqn\susyto{ \delta_{\bar\xi} \Phi_A^\dal   = \bar\xi_C\,
Q^C\, \calW_A^\dal |_{\bar \theta =0, \theta = M'}=
\bar\sigma^{mn}_{\dal\dbet} \bar\xi_A^\dbet \Phi_{mn} -   2\bar\xi_C
\sigma\cdot
\partial  \theta^C \, \Phi_A^\dal ,}
and the transformation of $\Phi_{mn}$ is given by
\eqn\susythr{\delta_{\bar\xi} \Phi_{mn}   = \bar\xi_C Q^C \Phi_{mn}
= -   2\bar\xi_C \sigma\cdot
\partial  \theta^C \, \Phi_{mn}.}
The $\bar\xi_C \sigma\cdot \partial \theta^C$ terms are proportional
to the momentum carried by the source and will not contribute to the
supersymmetry variation of any amplitude since the sum of the momenta
carried by the sources vanishes.  We may therefore drop the second
terms on the right-hand sides of \susyto\ and \susyxi.  The $\eta$
supersymmetry transformations act as shifts of $\Mp$, as can see from
the explicit expression,
\eqn\susyeta{\eqalign{
  \delta_{\eta} \Phi_{AB}& = \eta^C_\alpha \bar Q_C^\alpha
\calW_{AB}|_{\bar \theta =0, \theta = \Mp} = \eta^C_\alpha
{\partial\over \partial \Mp^C_\alpha}\, \Phi_{AB} .  \cr}} This
accounts for eight of the nonlinearly realized supersymmetries of the
collective coordinate action.

The expansions for the source superfields terminate with the terms of order
$\Mp^4$ upon
using the
 equations of motion and have the explicit form,
\eqn\supone{\eqalign{
\Phi_{mn}(x^n, \Mp) =& F^+_{mn} + i\Mp^{ A} \sigma_{[m} \, \partial_{n]}
\bar\Lambda_A +4 \Mp^B \sigma_{[m}^{\ \ p} \Mp^A \partial_{n]} \partial_p
\varphi_{AB}\cr
& + 2 \epsilon_{ABCD} \Mp^B \sigma_{p[m}\Mp^A \Mp^C
\partial_{n]} \partial_p \Lambda \cr
& + \epsilon_{ABCD}
\Mp^B\sigma_{p[m}\Mp^A\; \Mp^C \sigma^{kl}\Mp^D\; \partial_{n]}
\partial_p F^-_{kl},\cr }}
 \eqn\suptwo{\eqalign{ \Phi_A^{\dot
\alpha}(x^n, \Mp) = & \bar \Lambda_A^{\dot \alpha} -4i\Mp_\beta^B\,
\sigma^{n\, \beta\dot \alpha}\, \partial_n\varphi_{AB}-2i \epsilon_{ABCD}
\Mp_\beta^B\, \sigma^{n\, \beta\dot \alpha}\; \Mp^C \partial_n \Lambda^D
\cr
& +i \epsilon_{ABCD} \Mp_\beta^B\, \sigma^{n\, \beta\dot \alpha}
\;\Mp^C\sigma^{pq} \Mp^D \partial_n F^-_{pq} \cr
& + \epsilon_{ABCD}
\Mp_\beta^B\, \sigma^{n\, \beta\dot \alpha} \;\Mp^C\sigma^{pq} \Mp^D
\partial_n \partial_{[p}\Mp^E \sigma_{[q} \partial_n \partial_{p]}
\bar\Lambda_E ,\cr}}
 \eqn\supthree{\eqalign{\Phi_{AB}(x^n, \Mp) =& \varphi_{AB} +{1\over 2}
\epsilon_{ABCD}
 \Mp^{[C\alpha} \, \Lambda^{D]}_\alpha -{1\over 4} \epsilon_{ABCD}
\Mp^C\sigma^{mn} \Mp^D \,
 F^{-}_{mn} \cr & + {i\over 4} \epsilon_{ABCD} \Mp^C \sigma^{mn} \Mp^D
  \;
\Mp^{E} \sigma_{[m} \partial_{n]}\bar \Lambda_E \cr
  & + \epsilon_{ABCD}
\Mp^C\sigma^{mn}\Mp^D \,
 \Mp^E\sigma_{pm} \Mp^F \partial_n\partial^p \varphi_{EF}.\cr}}
The action \fullact\ only involves the relative superfields,
\eqn\relshat{ \hat\Phi_{AB} = \Phi_{AB} -\chi_{AB}, \qquad \hat
 \Phi_{A\dal} = \Phi_{A\dal} - \lambda_{A\dal}.}  Consequently, in the
presence of open-string sources the $\bar\xi$ transformation of $\mu$
in \transthree\ is replaced by the translationally-invariant
expression, \eqn\transtnew{
\delta_{\bar\xi} \mu^A = -4i w_{\dot \alpha} \bar{\hat \Phi}^{AB}\,
\bar\xi_B^{\dot \alpha},}
while $\delta_{\bar\xi} w$ remains unchanged.  The $\bar \xi$ variation of the
action \fullact\ is given by,

\eqn\varx{\eqalign{\delta_{\bar\xi}
S_{-1}[\Phi_{mn},\Phi^A_\dal,\Phi_{AB}]  = &-i\delta_{\bar\xi} \Phi_{mn} \bar
\eta^c_{mn} W^c + 2i \delta_{\bar\xi} \hat \Phi_{AB} \bar \mu^A \mu^B +
 i \delta_{\bar\xi} \hat \Phi_A^\dal (\bar \mu^A w_\dal + \mu^A\bar w_\dal)
\cr &
  + 4
\delta_{\bar\xi} \hat \Phi_{AB}\, \hat{\bar \Phi} {}^{AB}  \, W_0
 + (\half g_0^2 W_0 +2
\hat\Phi_{BC}^2) \, (\bar w \bar\xi_A\, \mu^A + w \bar\xi_A\,
\bar\mu^A) \cr & -i \Phi_{mn} (\bar w\bar \sigma^{mn} \bar\xi_A\, \mu^A -
\bar\mu^A\,\bar \xi_A \bar\sigma^{mn} w) \cr & + i\hat \Phi_{A\dal}
(\bar \mu^A \bar \xi_B^\dal \mu^B + \mu^A\bar
\xi_B^\dal \bar \mu^B
-4i \bar w^\dal \bar{\hat \Phi}^{AB}\, \bar \xi_B w -4iw^\dal
 \bar{\hat\Phi}^{AB}\, \bar \xi_B \bar w) \cr &
  + 8\hat \Phi_{AB} (\bar
 \mu^A\, w  \hat{\bar\Phi} {}^{BC}\bar \xi_C +\bar w
\hat{\bar\Phi} {}^{AC}\bar \xi_C\, \mu^B). \cr}}

  The requirement that $\delta_{\bar\xi}
S_{-1}[\Phi_{mn},\Phi^A_\dal,\Phi_{AB}]
=0$ determines the transformations of the D3-brane source superfields
which are given by
\eqn\varias{ \delta_{\bar\xi}\Phi_{mn} = 0, \qquad \delta_{\bar\xi} \hat
\Phi_A^\dal =
\bar\sigma^{mn}_{\dal\dbet} \bar\xi_A^\dbet\big( \Phi_{mn} +{i\over 2}
g_0^2 W_{mn}\big),\qquad
 \delta_{\bar\xi} \hat \Phi_{AB} = \bar\xi_{[A} \Phi_{B]} ,}
where the D3-brane
fields are evaluated at $x^m = a^m$.  These transformations differ
from the $\bar \xi$ transformations of the ${\cal N}=4$ theory only by
the term proportional to $g_0^2$.

\newsec{Integration over collective coordinates}

The effect of the D-instanton on open-string scattering amplitudes is
obtained by the integration over the collective coordinates,  which is
weighted by the instanton action \fullact\ with D3-brane source terms
included.
\eqn\intcoll{ Z[\Phi_{mn}, \Phi_A^{\dot \alpha},
\Phi_{AB}] = {\cal C} \int d^8M'\, d^8\lambda\, d^4\mu\, d^4 \bar{\mu}\,
d^4a\, d^6\chi\, d^2w\, d^2\bar w\, e^{-S_{-1}[\Phi_{mn}, \Phi_A^{\dot
\alpha},
\Phi_{AB}].}}
 The nomalization
${\cal C}$ will only be determined up to an overall numerical
constant, $c$,  although the  dependence  on $\alpha'$ and the string
coupling, $e^\phi$, is  important in the following and will be displayed.
This normalization has the form
\eqn\calcdef{{\cal C} = c\,  \times g_0^3 \times g_4^4,}
 where   $g_{p+1}$ is defined in \gpdef. The factor of $g_0^3$ comes from
integrating out
the auxiliary field $D_c$ and the factor of $g_4^4$ comes from
the normalization of the  instanton collective coordinate measure
\refs{\doreythree} for $N=1,k=1$.

\subsec{Integration over fermionic collective coordinates}

Integration over the fermionic  coordinates only gives a non-zero
 result when there is  an appropriate number of insertions of
 fermionic sources.  We will consider the $\mu$
 integrations first.  These can be performed in several ways, giving
rise to distinct sectors according
 to how many factors  of $\lambda$ are brought down from
the expansion of $e^{-S_{-1}}$.
   For  example,  one way to
 perform the  $\mu$  integrations is to bring down
four factors of $ \bmu^A w_\dal \lambda_A^\dal$ and four factors
of $ \bar w_\dal \mu^A \lambda_A^\dal$ from  $e^{-S_{-1}}$.
In that case the combined $\mu$,
$\lambda$ integrations enter in the form,
\eqn\fermulam{\int d^8\lambda d^4\mu d^4 \bar{\mu} \exp\Big( i\bmu^Aw_\dal
\lambda_A^\dal+i  \bar w_\dal \mu^A \lambda_A^\dal\Big)= (\bar{w}^\dal
w_\dal)^4= W_0^4.}
We will call this sector, in which all eight powers of $\lambda$
are soaked up by the $\mu$ integration,   the `minimal' sector.  More
 generally, there
are sectors in which  factors of $ (\chi_{AB}-\Phi_{AB})\bar
\mu^A\mu^B$ or $\Phi_A^{\dot \alpha} \, \bar \mu^A w_{\dot \alpha}$
are brought down from the expansion of $e^{-S_{-1}[\Phi_{mn},
\Phi_{\dot \alpha}^A\Phi_{AB}]}$.
  In such cases
the integration over $\mu$    brings down  less than eight powers
of $\lambda$, resulting in  $2$, $4$,
$6$ or $8$ unsaturated components of $\lambda$.
Since there are no  open-string D3-brane sources that couple to
$\lambda$, these
sectors of the integral will vanish in the absence of  closed-string
 sources.   We will see later that
closed-string sources  couple to a  disk with a number of $\lambda$
vertex operators
(and a pair of twist operators) attached so these can provide the
 missing powers of $\lambda$.    However, it will  still be
true that the instanton
induced  interactions  of lowest
dimension  arise in the  minimal sector.

For the moment we will consider  case  in which there are no
closed-string sources so that only the minimal sector is relevant.
  The integration over  $M'$ necessarily
brings down powers of the open-string sources since $M'$ does not appear
in the source-free action, $S_{-1}[0,0,0]$ \fullact.  The non-zero
instanton-induced
amplitudes that come from expanding   \intcoll\ in powers of the
sources   have the form  (absorbing an overall constant into $c$),
\eqn\nonzw{A^{inst}_{\Phi_1\dots\Phi_r}={\cal C}\int d^4 a d^2w d^2\bar w
d^6\chi\,
d^8 M'\, W_0^4\,e^{-\hat S_{-1}} \,
\langle \Phi_1\rangle_{w\bar w;m_1} \dots
 \langle \Phi_r\rangle_{w\bar w;m_r},}
where $\sum_r m_r =8$ and $\Phi_r$ are the component D3-brane fields that
couple to $m_r$
$V(M')$ vertex operators and two bosonic twist operators.  The  integration
over
the $a^m$ coordinates leads to an overall momentum conservation delta function,
 $\delta^{(4)}(\sum_r k_r^m)$, which will be suppressed in the following
formulae.  The quantity
 $\hat S_{-1}$ in \nonzw\ is given by
\eqn\shatdef{\hat S_{-1}  = -2\pi i \tau +{1\over 4}g_0^2 W_0^2
+ \chi_a^2  W_0.}
 In order to perform the $w$ and $\bar w$
integrations it will be  convenient  to change variables to  $W^c$
defined in \wdefs. The integration measure for $w_\dal$ becomes
\eqn\measure{\int d^2 w d^2 \bar w = 2\pi \int {d W^1dW^2dW^3\over
W_0},}
where $W_0^2 = \sum_{c=1}^3 (W^c)^2$.  Since $W^c$ appears
quadratically in the action
the $W^c$ integral is a simple Gaussian.
The
 integration over the four bosonic moduli $a^n$, which  imposes
conservation of overall
longitudinal momentum on the scattering amplitudes, has been performed in
 writing \nonzw.

For definiteness we will now  consider the case of four $\langle
\varphi\rangle_{w\bar w;2}$  insertions in detail. In that case the
 collective coordinate integration  is given by
\eqn\collint{ \eqalign{ A^{inst}_{\varphi^4}&={\cal C}\int {d^3W^c\over
W_0}
 d^6 \chi d^8 M'\,e^{-\hat S_{-1}} \, \langle
\varphi\rangle_{w\bar w;2}\,
  \langle \varphi\rangle_{w\bar w;2}\,
\langle \varphi\rangle_{w\bar w;2}\,
\langle \varphi\rangle_{w\bar w;2}\cr
 & = {\cal C}\int {d^3
W^c\over W_0} d^6 \chi d^8 M'\,e^{-\hat S_{-1}}
   \bar \eta^{d_1}_{m_1n_1}
\bar \eta^{d_2}_{m_2n_2}
\bar \eta^{d_3}_{m_3n_3}
\bar \eta^{d_4}_{m_4n_4}W^{d_1} W^{d_2} W^{d_3}  W^{d_4}\cr
&\times
\eta^{c_1}_{p_1m_1}\, \eta^{c_2}_{p_2m_2}\, \eta^{c_3}_{p_3m_3}\,
 \eta^{c_4}_{p_4m_4}
  M' \Sigma^{a_1} \tau^{c_1} M'  \partial_{n_1}
\partial^{p_1} \varphi^{a_1} (x_1)
 \  M' \Sigma^{a_2} \tau^{c_2} M'  \partial_{n_2}
\partial^{p_2}
\varphi^{a_2} (x_2)\cr
 &\times\  M' \Sigma^{a_3} \tau^{c_3} M'  \partial_{n_3}
\partial^{p_3} \varphi^{a_3} (x_3)
  M' \Sigma^{a_4} \tau^{c_4} M'  \partial_{n_4}
\partial^{p_4} \varphi^{a_4} (x_4)
 ,  \cr }}
where we have used the relation between $\sigma^{mn}$ and $\tau^c$ in
appendix A to write
\eqn\rewrit{{M'}^{A}\sigma_{pm} {M'}^{B}
\varphi_{AB}(x)=  M' \Sigma^{a} \tau^{c} M'\, \varphi^a(x) \eta^c_{pm}.}
It is convenient to work in momentum space so that $\partial^n  = ik^n$ in
 \collint, where the integration over   $a_n$ in
\intcoll\ guarantees  momentum conservation, $\sum_r k_r^n =0$.

In contemplating the  $W^c$ integration it is useful to transform to polar
coordinates $(W_0, \theta,\phi)$, where the measure is
\eqn\wmeas{d^3 W^c W_0^{-1} =  dW_0\, W_0\,  \sin \theta d\theta d\phi,}
and the  angular part of the integral \collint\  gives,
\eqn\winto{\int   \sin\theta d\theta\, d\phi\,
W^{d_1}W^{d_2}W^{d_3}W^{d_4}  = {4\pi \over 15}
W_0^4
\,\Big(\delta^{d_1d_2}\delta^{d_3d_4}+\delta^{d_1d_3}\delta^{d_2d_4}+
\delta^{d_1d_4}\delta^{d_2d_3}
\Big) ,}
where the radial $W_0$ variable will be integrated later.

In order to simplify the structure of the integrand we may make use of
the identities involving products of  'tHooft's $\eta$ symbols which
follow   from their definition (see appendix A).  A very simple way
to
evaluate the fermionic integrations is given by choosing a particular
frame for the momenta in
which one component, $k_r^0$, of each momentum vanishes.  In that
case, after using  the mass-shell condition ($k_r^2=0$) we have the
equation,
\eqn\etaids{\eta_{m_r p_r}^{c_r} \bar \eta_{m_r n_r}^{d_r}
k_r^{p_r} k_r^{n_r} = 2 k_r^{c_r} k_r^{d_r},}
Using this expression in \collint\ together with the
\winto\  leads to an expression for  $M'$ proportional to
\eqn\mprii{(s^2+t^2+u^2)  \int d^8 M'\left(
\prod_{r=1}^4    M^{\prime}\Sigma^{a_r} \tau^{c_r}
M^{\prime }  k_r^{c_r} \varphi^{a_r}(x_r)  \right)  ,}
  The factor of $(s^2+t^2+u^2)$  comes from the contractions of the
  momenta with the $\delta^{d_rd_s}$ factors in \winto, which arose
  from the $W^c$ integrations.

The integrand  of \mprii\  involves the fermion bilinears
 $ M^{\prime}\Sigma^{a_r} \tau^{c_r}
M^{\prime }$ (where $a_r=1,\cdots,5$ and $c_r=1,\cdots,3$) which  have
the same structure as the SO(8) spinor  bilinears  of light-cone
superstring theory.
A standard argument relates the integral over a product of
the components of a  fermionic $SO(8)$ spinor to the tensor $t_8$ in
 \kininv\ so that
the  integration over $M'$ produces
\eqn\mints{\eqalign{\int d^8 M'\left( \prod_{r=1}^4    M' \Sigma^{a_r}
\tau^{c_r} M'\,  k_r^{c_r} \varphi^{a_r}(x_r) \right) &  =
t_8^{a_1c_1 \dots a_4c_4}  k_1^{c_1} \varphi^{a_1}(x_1) \dots
k_4^{c_4} \varphi^{a_4}(x_r) \cr
&=tu\ \varphi(x_1)\cdot \varphi(x_2)\ \varphi(x_3)\cdot \varphi(x_4) +
\ {\rm perms} .\cr}}
Hence the instanton induced $(\partial^2 \varphi)^4$ term is given by
\eqn\instphires{A^{inst}_{\varphi^4}=(s^2+t^2+u^2)\big(tu
\varphi_1\cdot\varphi_2 \varphi_3\cdot\varphi_4 +su
\varphi_1\cdot\varphi_4 \varphi_2\cdot\varphi_3+st
\varphi_1\cdot\varphi_3 \varphi_2\cdot\varphi_4 \big)\; I_4,}
where $I_4$ denotes the remaining integrations over the
 bosonic collective coordinates, $\chi^a$ and $W^c$, which will be
 discussed in the next
 subsection. Note that  \instphires\ has  the same form as the kinematic
dependence of the tree-level diagram in \kinfac.  Although our derivation
of this
expression used  a particular frame, a more careful
analysis gives the same result, as is expected from Lorentz
invariance.

In a similar manner one can discuss amplitudes for scattering any of
the other open-string states.  For example,
the insertion of  two $\langle F^+\rangle_{w\bar w;4}$ sources saturates
the $M'$
integration, leading to a possible   $\partial^2 F^+\partial^2
F^+$ term in the action. However this is a total derivative and vanishes
after integration.  At least two
powers of the  source $\langle F^-\rangle_{w\bar w;0}$ have to be brought
down from
the exponential to   produce a non vanishing
result,
\eqn\collinta{ \eqalign{ A^{inst}_{F^4} = & {\cal C}\int {d^3W^c\over W_0} d^6
\chi d^8 M'\, e^{-\hat S_{-1}}\, \langle
F^+\rangle_{w\bar w;0}\,
  \langle F^+\rangle_{w\bar w;0}\,
\langle F^-\rangle_{w\bar w;4}\,
\langle F^-\rangle_{w\bar w;4}\cr
=&  {\cal C} \int {d^3
W^c\over W_0} d^6 \chi d^8 M'\,e^{- \hat S_{-1}}
   \bar \eta^{d_1}_{m_1n_1}
\bar \eta^{d_2}_{m_2n_2}
\bar \eta^{d_3}_{m_3n_3}
\bar \eta^{d_4}_{m_4n_4} \cr &
\times W^{d_1} W^{d_2} W^{d_3}  W^{d_4}
 F^+_{m_1n_1}(x_1)F^+_{m_2n_2}(x_2) \cr
& \times \epsilon_{A_3B_3C_3D_3}
{M'}^{A_3}\sigma_{p_3m_3}{M'}^{B_3}{M'}^{C_3}
\sigma^{k_3l_3}{M'}^{D_3}\;
\partial_{n_3} \partial_{p_3} F^-_{k_3l_3}(x_3)\cr
& \times \epsilon_{A_4B_4C_4D_4}
{M'}^{A_4}\sigma_{p_4m_4}{M'}^{B_4}{M'}^{C_4} \sigma^{k_4l_4}{M'}^{D_4}\;
\partial_{n_4} \partial_{p_4} F^-_{k_4l_4}(x_4).}}
The integration over $M'$ and $W^c$  produces (after
integration by parts) a term of the form
\eqn\finter{s^2( F^+)^2(  F^-)^2,}
The full amplitude is given by summing over the two other choices for the
location of the fermionic zero modes. Since $( F^+)^2(  F^-)^2$
 is invariant under permutations of the fields the result is
\eqn\finterb{A^{inst}_{(F^+)^2(F^-)^2}=(s^2+t^2+u^2) (F^+)^2(F^-)^2
=(s^2+t^2+u^2) \big(F^4 - {1\over 4} (F^2)^2\big)\; I_4, }
which is the same  tensor structure as the perturbative contribution
\ordertwo. A straightforward extension of this argument also gives the
 instanton-induced interaction between two scalars and two
field strengths as $\partial F^- \partial F^+ (\partial^2 \varphi)^2$
which  also agrees with the structure of the analogous tree-level term
\ordertwo.

\subsec{Integration over bosonic collective coordinates}

In general we  want to consider processes with $r$ external
D3-brane open-string
fields  in \nonzw, where the case of $r=4$ was considered in
the previous subsection
and we will   consider $r=6,8$ in section 6.
In order to accommodate all these
 cases we can rewrite the generating function \intcoll\ after
performing the
 $\lambda$, $\mu$ and $\bar \mu$ integrals and  in the
absence of  closed-string sources,\foot{We are still considering the
`minimal' case defined in the first paragraph of 5.2,
which is the only sector of relevance in the absence
of closed-string sources.}  in the form,
\eqn\formamp{Z[\Phi_{mn}] = {\cal C} \int d^4a^m\,
d^6 \chi^a \, d^8 M'\,  {d^3W^c\over W_0}\,e^{-\hat S_{-1}}\,
e^{- i \Phi_{mn}(M')W^{mn}}.}
This generates all the instanton-induced open-string amplitudes,
 which are obtained by expanding the exponential   to extract the terms that
are eighth order in $M'$.
The  modular invariant interactions (such as $(\partial F)^4$)
 come from the term in which four
powers of $W^{mn}$ are brought down from the exponential.   As will be
discussed in more detail in section 6, there are also
interactions that transform with modular weight
$(1,-1)$  (such as $(\partial^2 \varphi)^2 (\partial \bar
\Lambda)^4$)  that arise from terms with
six powers of $W^{mn}$ as well as interactions
 that transform with weight
$(2,-2)$ (such as $  (\partial \bar
\Lambda)^8$) that come from terms with eight powers of $W^{mn}$.
It is evident by a simple rescaling of $W$ that for fixed $\chi$ the
perturbative expansion
of \formamp\ is actually an expansion in powers of $g_0^2 L^{-4}$, where $L
 =|\chi^a|$.  This
corresponds to the genus expansion of the world-sheet configurations
that contribute to the process.  Every term in this series is singular
in the $L=0$ limit although the exact expression \formamp\ is
well-defined.

In order to evaluate the amplitudes for scattering D3-brane fields
it is instructive to perform the $\chi$ and $W^c$   integrations in
\formamp\ before
performing the $M'$ integrations.    Transforming the $W^c$ integration
to polar coordinates and performing the gaussian $\chi^a$ integrals
results in the expression,
\eqn\znewdef{Z[\Phi_{mn}] = (2\pi)^5 {\cal C} \int d^4 a\, dW_0\,  \sin\theta
d\theta\, W_0^2\, \exp\left(- i W_0\, \cos\theta\, |\Phi|  -
{g_0^2\over 4} W_0^2\right),}
where $|\Phi| = (\Phi_{mn}\Phi^{mn})^{1/2}$.
After changing variables from $W_0$ to $\hat W_0 = g_0W_0/2$
and performing the $\theta$ integration the result is
(again absorbing an overall constant into $c$)
\eqn\zagain{Z[\Phi_{mn}] =
 c  g_4^4\int d^8 M'  \int d^4 a^m d\hat W_0\, \hat
W_0^2\, {\sin y\over y}\, e^{- \hat W_0^2} \equiv c
\int d^8 M' \,  \sum_{r=even}\,
I_r\,  |\Phi|^r, }
where the $da^n$ integral gives a suppressed factor of
$\delta^{(4)}(\sum_{s=1}^r k_s^m)$ and
\eqn\ydef{y = g_0^{-1}\, \hat W_0\, |\Phi|.}
The various terms that enter into the expansion of the integrand to eighth order in $M'$ are easily extracted from \zagain\ by using the expansion
\eqn\sinnexp{{\sin y\over y} =
\sum_{r=even}^\infty  {y^r \over (r+1)!}(-1)^{r/2}.}
The  $\hat W_0$ integration in \zagain\ is contained in
\eqn\ifun{\eqalign{I(e^\phi) & = g_4^4 c \int d\hat W_0\,
\hat W_0^2 \,
e^{-\hat W_0^2} {\sin y\over y} \cr
&  =  \sum_{r=even}{\alpha'}^r \tau_2^{-2 + r/2}\, I_r,\cr}}
where the coefficients $I_r$ are given by
\eqn\resi{I_r = c\; 2^{4+r}\pi^{2+3r/2}
{\Gamma\left({r\over 2} + {3\over 2}\right)\over
\Gamma(r+2)}(-1)^{r/2}.}
The gaussian $\hat W_0$ integration in each term in the expansion of
the integrand
of   \zagain\  peaks at $\hat W_0 =0$
and $\langle \hat W_0 \rangle =0$.  However, if a
non-zero background $B_{mn}$ field is present the integrand  in  \formamp\
is  multiplied by $e^{- iB_{mn}W^{mn}/2\pi \alpha'}$ so that only
the gauge invariant combination $B_{mn} + 2\pi \alpha' F_{mn}$ enters.
In that case  the $W_0$ integration
 peaks   at $W_0 \sim |B_{mn}|/g_0^2\pi
\alpha'$.
The systematics of the induced interactions
is then quite different since there are odd powers of $y$ in the
expansion  in \ifun.
Furthermore, as discussed in \seibergwitten, a finite $B$
background rotates the boundary conditions and changes the
identification  of
the broken and unbroken supersymmetries.  The case of non-zero $B$
will not be discussed in this paper.

  We see from \zagain\ that the leading contributions to the D-instanton
induced interactions   in the  $r=4$ terms  are
independent of $\tau_2 = e^{-\phi}$ and their dependence on the
string length is given by
${\alpha'}^4$.   We also saw  in section 5.1 that the
  ${M'}^8$ terms in the expansion of
 $|\Phi_{mn}|^4$  have the same tensor structure as the
corresponding tree-level terms.  These facts mean that we can identify the
  leading instanton contribution to the $r=4$ terms with the
instanton  terms in the weak-coupling expansion of
\fourfmod, using \expanh\  (which  has the property that the
leading instanton
contributions are  actually exact). A more thorough treatment would
also check the absolute normalization of the instanton contribution
which we have not determined.
The dependence of \intcoll\  on $\alpha'$ and
 $\tau_2$ for terms with   $r=6$ is given by
${\alpha'}^6 \tau_2$ and  for the $r=8$ terms by ${\alpha'}^8 \tau_2^2$, which
 will be important  for other open string interactions discussed in
 section 6.

\newsec{Other open-string interactions}

The interactions described in section 5 are ones in which the
coupling constant appears
in the modular invariant function $h(\tau,\bar\tau)$ defined in
\modfun.  More generally,
there are interactions that transform with non-zero modular weight.
The modular invariance of the D3-brane theory is best examined in the Einstein frame
of the bulk theory with
metric $g^{(E)} =\tau_2^{1/2}g^{(s)}$ or $e^{(E)} = \tau_2^{1/4} e^{(s)}$
(where the superscripts ${}^{(E)}$ and ${}^{(s)}$
 denote the Einstein and string frame, respectively and
$e \equiv e_{\mu\, m}$ is the vierbein),
 since the metric is $SL(2,Z)$ invariant in that frame.  The
conventional normalization of the ${\cal N}=4$ Maxwell action is
obtained in this frame
provided the  Einstein-frame fields are related to those in the string frame in the
following manner,
\eqn\streins{A_m^{(s)} = A_m^{(E)}, \qquad \Lambda^{(s)} =
  \tau_2^{-1/8} \Lambda^{(E)},
\qquad \bar\Lambda^{(s)} = \tau_2^{-1/8} \bar\Lambda^{(E)}, \qquad
\varphi_{AB}^{(s)}  = \tau_2^{-1/4} \varphi_{AB}^{(E)}.}
 In  the
following the superscript  ${}^{(E)}$ will often be dropped.  With
these field normalizations a factor of $\tau_2$ multiplies only
the Maxwell term in the free action.

It will be convenient to absorb a power of $e^\phi$  into the  field strengths so that  (from \aselftrans)
the combination $\hat F^\pm = \tau_2^{1/2} F^\pm$ transforms with a phase under
$SL(2,Z)$,
\eqn\bself{\hat F^\pm  \to \left({c \bar\tau + d \over c \tau +
      d}\right)^{\pm 1/2}\,
\hat F^\pm .}
The fermion fields also transform with a phase,
\eqn\lamphase{\bar \Lambda \to \left({c \bar\tau + d \over c \tau + d}\right)^{1/4}
 \bar\Lambda, \qquad   \Lambda \to \left({c \bar\tau + d \over c \tau
     + d}\right)^{-1/4}
 \Lambda,}
 while the scalar field $\varphi_{AB}$ is inert.\foot{ These
transformation rules can be obtained by considering $N=4$ Maxwell theory
coupled to $N=4$ supergravity \refs{\deroo,\bergsh}. }

\subsec{Instanton induced interactions}

The tree-level open-string amplitude with external fermions is a well-known
 generalization of
\treeone\ that gives rise to higher derivative interactions
of the form,
\eqn\highfera{{\alpha'}^4\int d^4x\,  \det \, e^{(s)}
\tau_2(\partial^2\Lambda^{(s)})^2(\partial\bar\Lambda^{(s)})^2 ={\alpha'}^4
 \int d^4x\, \det \, e\, \tau_2(\partial^2\Lambda)^2(\partial\bar\Lambda)^2,
 }
 where the precise contractions between the indices are  defined by
 the kinematic factor of the four-fermion amplitude
 \refs{\greenschwarza}.  Similarly, there are terms of the same
 dimension of the   schematic form $\partial \hat F^+ \partial \hat F^-
 \partial\bar \Lambda\partial^2
 \Lambda$ and $(\partial^2 \varphi)^2\partial\bar \Lambda\partial^2
 \Lambda$, which are also
 modular invariant.  It seems plausible that the  coupling constant dependence
 again enters
via  the modular function $h(\tau,\bar\tau)$ that is the coefficient
 of the $(\partial \hat F)^4$
term.  This is supported by the discussion  in section 5.3 where the
instanton contributions to these terms  were seen to be related by
 ${\cal N}=4$  supersymmetry transformations since they were obtained from
a supersymmetric generating function.

Similarly, there are terms that transform with non-zero modular
weights.  Although the
tree-level expressions for these interactions are complicated to
analyze it is easy to
deduce their presence from the instanton induced interactions.  For
example, there is an
eight-fermion term which has a tree-level contribution,
\eqn\highferc{{\alpha'}^8\int d^4x\, \det \, e^{(s)}\, \tau_2 (\partial \bar
 \Lambda^{(s)})^8
= {\alpha'}^8\int d^4x\, \det \, e\, \tau_2 (\partial\bar \Lambda)^8.}
Since $\bar\Lambda^8$ transforms with a phase $(c\tau+d)^2/
(c\bar\tau+d)^2$ the modular-invariant interaction must have the form,
\eqn\modlam{  {\alpha'}^8\int d^4x\, \det \, e\, h^{(2,-2)}(\tau,\bar\tau)\,
(\partial\bar \Lambda)^8,}
where the modular form $h^{(2,-2)}(\tau,\bar\tau)$ transforms with a
compensating
 phase $(c\bar \tau+d)^2/ (c\tau+d)^2$ and has a large $\tau_2$ (small
coupling constant)
 expansion that starts with the tree-level term.
In a similar manner there are interactions of the form
\eqn\moreint{{\alpha'}^6\int d^4x\, \det \, e
\,h^{(1,-1)}(\partial\bar \Lambda)^4\,
(\partial^2\varphi)^2, \qquad  {\alpha'}^6\int d^4x\, \det \, e\ h^{(1,-1)}
 (\partial\bar \Lambda)^4  \hat F^+
\partial^2 \hat F^-,
\qquad \dots.}

This structure is similar to the structure of higher derivative
terms in the bulk type IIB action which are of the same dimension
as $R^4$ \refs{\greengutkwon}. In that case there are interactions with
integer weights $(w,-w)$ with $0\le w \le 12$ while in the present
case the weights span the range $0\le w \le 2$.  Following the same
path as in that case, it is
 reasonable to conjecture that the relations between the modular forms
$h^{(w,-w)}$
are  obtained by applying successive modular covariant derivatives
 on the
modular function,\foot{The notation $h^{(0,0)}\equiv h$ is suited to
the the generalization to modular forms of non-zero weight.}
$h^{(0,0)}(\tau,\bar\tau)\equiv  h(\tau,\bar\tau) = \ln |\tau_2\eta(\tau)^4|$.
Thus,
\eqn\modfa{\eqalign{h^{(1,-1)} & = D_0 h^{(0,0)}(\tau,\bar\tau)=
i\tau_2 {\partial\over \partial\tau}
 h^{(0,0)} =
 - {\pi\over 6}\tau_2 \hat E_2 \cr &
  = -{\pi\over 6}\tau_2 +{1\over 2} +
4\pi \tau_2 \sum_{N=1}^\infty\sum_{m|N} me^{2i\pi\tau N}  ,\cr}}
where $E_2$ is the second  Eisenstein series (which is holomorphic but
not modular covariant) while  $\hat E_2$ is the non-holomorphic
Eisenstein series of weight $(2,0)$.
\eqn\etwodef{\hat E_2 = E_2 -{3\over \pi \tau_2}, \qquad E_2=
 1-24\sum_{n=1}^\infty {nq^n\over 1- q^n}, }
 where $q= e^{2\pi i \tau}$.
 More generally, the  covariant
derivative acting on a modular form $h^{(w,-w)}$ is
$D_w = i(\tau_2\partial/\partial \tau - iw/2)$ converts it to a form
$h^{(w+1,-w-1)}$.  Thus, applying a covariant
derivative to $h^{(1,-1)}$ gives
\eqn\modfb{\eqalign{h^{(2,-2)}& = D_1 h^{(1,-1)}= i\big(\tau_2{\partial\over
\partial\tau}-{i\over2}\big)
h^{(1,-1)}\cr
&= - {\pi\over 6}\tau_2^2 \big( i{\partial\over \partial\tau}+
{1\over \tau_2}\big)\hat E_2\cr
&=-{\pi\over 36}\tau_2^2\big(E_4-\hat E_2^2\big)}}
where we have used the fact that
\eqn\deretwo{\big( i{\partial\over \partial\tau}+
{1\over \tau_2}\big)\hat E_2= {\pi\over 6}(E_4-\hat E_2^2)}
 and the fourth Eisenstein  series is  defined by
\eqn\eisenfor{E_4 = 1+240 \sum {n^3 q^n\over 1- q^n}.}
 The instanton expansion of $h^{(2,-2)}$ is given by
\eqn\intwo{ h^{(2,-2)}=   -{\pi\over 6}\tau_2 +{1\over 4}+
 4\pi \tau_2
\sum_{N=1}^\infty\sum_{m|N} m e^{2\pi i \tau N}- 8\pi \tau_2^2
\sum_{N=1}^\infty\sum_{m|N} mN e^{2\pi i \tau N}.}

All the expressions $h^{(w,-w)}$ possess tree-level and one-loop
 terms
that correspond to perturbative effects in the world-volume of the
D3-brane.  The instanton contributions  to the expansions of
$h^{(0,0)}$ in \expanh,  $h^{(1,-1)}$ in
\modfa\ and   $h^{(2,-2)}$ in  \intwo\
 are multiplied by different powers of $\tau_2$. The leading terms in
the  weak coupling limit ($\tau_2 = g^{-1} \to \infty$) have powers
$\tau_2^0$, $\tau_2$ and $\tau_2^2$, respectively. These powers
correspond to the terms with $r=4$, $r=6$ and $r=8$ in \ifun, which
summarizes the results of the explicit leading-order D-instanton
calculations,
which automatically take account of supersymmetry constraints.
We therefore have some evidence
that the conjectured form $SL(2,Z)$-covariant prefactors $h^{(1,-1)}$
and $h^{(2,-2)}$ are given (up to overall constant normalizations) by
\modfa\ and \modfb, respectively.

 In  \refs{\bainbachasgreen}
the relation of the D3 brane to
the M5 brane wrapped on a two-torus was used to
show that  $h^{(0,0)}$ can be obtained  from a one loop
calculation in the world-volume theory of the M5 brane. It would
be   interesting to use an analogous one-loop calculation to
reproduce the conjectured form of $h^{(1,-1)}$ and $h^{(2,-2)}$.

\newsec{Closed-string interactions}

The effective action of a D3-brane depends in a crucial manner on the
geometry that describes the embedding of the world-volume in the target
space-time.  This means that in addition to the dependence on the
fields in the  Yang--Mills supermultiplet it is important to
understand terms in  the effective action that involve the
gravitational sector pulled back to the world-volume.  These have been
discussed to some extent in the literature.  For example, terms of the
general form $R^2$ arising from  tree-level scattering of a graviton
from a Dp-brane were obtained in
\refs{\myers,\garmyers,\hashkleb}
where the
two-point tree-level graviton amplitude was written as
\eqn\twopt{\eqalign{
A^{tree}_{R^2}(h^{(1)},p_1; h^{(2)},p_2) &= - {1\over 8}T_{(p)}
\;\alpha^{\prime 2}\; K(1,2)\;
{\Gamma{(-\alpha^\prime t/4)}\Gamma{(\alpha^\prime q^2)}\over
\Gamma{(1-\alpha^\prime t/4+\alpha^\prime q^2)}} \cr
   & =  {1\over 2}T_{(p)}\;K(1,2)\;
\left({1\over q^2 t} + {\pi^2\alpha^{\prime 2}\over 24}
+ o(\alpha^{\prime 4})\right).}}
Here $h^{(1)}_{\mu\nu}$ and $h^{(2)}_{\mu\nu}$ $(\mu=0,\dots,9$)
are the polarization vectors
for the on-shell gravitons,
$q^2=p_1\cdot D\cdot p_1/2$ is the square momentum
flowing along  the
world-volume of the   Dp-brane and $t=-2p_1\cdot p_2$ is
the momentum transfer in the transverse directions and the matrix $D$
is the diagonal matrix
\eqn\ddef{D_\mu^{\nu}  =\left\{\eqalign{ \delta_\mu^{\ \nu} &\quad\quad
(\mu,\nu = 0,\dots, p),\cr
 - \delta_\mu^{ \nu}& \quad\quad (\mu,\nu = p+1,\dots, 9).}\right.}
 The kinematic factor $K$ is expressed in terms of the tensor $t_8$
which appear in
the four open-string scattering amplitude,
\eqn\kinematicfacb{K(1,2)=  t_8^{\mu_1\nu_1\rho_1\lambda_1 \mu_2\nu_2\rho_2\lambda_2 }
 \;D^{\bar \rho_1}_{\rho_1} D^{\bar
\lambda_1}_{\lambda_1} \;D^{\bar \rho_2}_{\rho_2} D^{\bar
\lambda_2}_{\lambda_2}\;
h^{(1)}_{\mu_1 \bar\rho_1}k^1_{\nu_1}
k^1_{\bar \lambda_1}h^{(2)}_{\mu_2 \bar\rho_2}k^2_{\nu_2}
k^2_{\bar \lambda_2}
  .}

Using the definition \ddef\ and the momentum invariants $q^2$ and
$t$ the kinematic factor \kinematicfacb\ can be written as
\eqn\kinematicfac{K(1,2)=\left(2q^2\,a_1+{t\over 2}\,a_2\right),}
with
\eqn\aoneatwo{\eqalign{
a_1&=\tr(h^{(1)}\cdot D)\,p_1\cdot h^{(2)}\cdot
p_1-p_1\cdot h^{(2)}\cdot
D\cdot h^{(1)}\cdot p_2-p_1\cdot h^{(2)}\cdot
h^{(1)}\cdot D \cdot p_1\cr
&\-p_1\cdot h^{(2)}\cdot h^{(1)}\cdot D
\cdot p_1-
p_1\cdot h^{(2)} \cdot h^{(1)}\cdot p_2+q^2\,\tr(h^{(1)}\cdot h^{(2)})
+\Big\{1\leftrightarrow2\Big\}\cr
a_2&=\tr(h^{(1)}\cdot D)\,(p_1 \cdot h^{(2)}\cdot
D \cdot p_2+p_2\cdot
D\cdot h^{(2)}\cdot p_1+p_2 \cdot D\cdot
h^{(2)}\cdot
D \cdot p_2)\cr
&+p_1\cdot D \cdot h^{(1)}\cdot D\cdot
h^{(2)}\cdot D
\cdot p_2-p_2\cdot
D\cdot h^{(2)}\cdot h^{(1)}\cdot D\cdot p_1
+q^2\,\tr(h^{(1)}\cdot D\cdot h^{(2)}\cdot D)
\cr&-q^2\,\tr(h^{(1)}\cdot h^{(2)})
-\tr(h^{(1)}\cdot D){\tr}(h^{(2)}\cdot
D)\,(q^2-t/4)
+\Big\{1\leftrightarrow2\Big\} .\cr }}
The form of this amplitude, together with some mild topological
input, leads to an explicit expression for the tree-level contribution
to the CP even  $($curvature$)^2$ terms in the action \refs{\bainbachasgreen},
\eqn\cpeven{\eqalign{
{\cal L}_{CP-even}^{(p)}& = {\tau_2\over 3\times 2^7 \pi}
\Big( ({ R}_T)_{\alpha \beta \gamma \delta} ({ R}_T)^{\alpha \beta \gamma
\delta} - \cr
&   - 2 ({ R}_T)_{\alpha \beta} ({ R}_T)^{\alpha \beta} -
{(R_N)}_{ \alpha \beta a b}{(R_N)}^{ \alpha \beta a b }
+ 2 \bar{R}_{ab}\bar{R}^{ab} \Big) . \cr}}
The notation in this expression is reviewed in appendix B.
As in the case of the open-string interactions discussed earlier  the
expressions in  brackets are modular invariant and a $SL(2,Z)$-invariant
action is obtained when  the factor of  $\tau_2$ is
replaced by a modular function.
The CP-odd  curvature terms in the D$p$-brane action are determined
by an anomaly-cancelling argument.

General arguments were given in \refs{\bainbachasgreen} that the modular
function  describing the exact scalar field dependence
of the non-perturbative contributions to the $($curvature$)^2$ terms
involving the tangent bundle (including the terms in the CP-odd part of the
action) is the function $h^{(0,0)}$ defined by \modfun.  However,
 little was said concerning
non-perturbative normal bundle  effects which is our main interest
in this section.  The dependence of the
tangential  and longitudinal
pull-backs of the curvature ($R_T$ and $R_N$) as
well as $\bar R$  on the second fundamental
 form leads to a dependence on the scalar
fields $\varphi_{AB}$. Thus, the terms  $\Omega^a_{mp}\Omega^b_{nq}-
\Omega^a_{mq}\Omega^b_{np}$ in $R_T$ and $\Omega^a_{mp}\Omega^b_{nq}
- \Omega^b_{mp} \Omega^a_{nq}$ in $R_N$ together with the terms bilinear in $\Omega$ that
enter into $\bar R$  give terms in \cpeven\ that are of the form $(\partial^2\varphi)^4$
and indeed correspond to the open-string scalar field interactions that we discussed in
the earlier sections \foot{ We are grateful to P. Bain for a useful
correspondence on this point.}.

We are now in a position to say more by generalizing the arguments of
the preceding sections to the situation in which there are
closed-string fields that couple via closed-string vertex operators.
There are three kinds of contributions
corresponding to the three possible boundary conditions on a disk: (a) The disk
with D3-brane conditions gives the standard tree-level gravitational contributions to the
Born--Infeld action; (b) The disk with D-instanton boundary conditions determines the
zero modes of the gravitational fields in the D-instanton background and will give
rise to $($curvature$)^2$ terms in the effective action; (c) The gravitational
coupling to a disk with $N$ and $D$ conditions (and two twist operators) -- these
effects are of relevance to the situation in which there is a non-zero background $B_{mn}$
field which will be commented on in the conclusion.  We will here give a detailed
discussion of  case (b) only.

\subsec{$($Curvature$)^2$ and related terms}

In the absence of a D3-brane the one-point
function of  any  field  in the IIB supergravity multiplet
in a D-instanton background may be obtained by coupling the field to a disk with Dirichlet
boundary conditions in all directions \refs{\greenguteffects}.  This multiplet of one-point
functions is generated by applying the sixteen broken supersymmetries, $\theta^{\cal A}
Q_{\cal A}$,  to the dilaton one-point function, where $\theta^{\cal A}$
(${\cal A} = 1,\dots,16$) is a Weyl Grassmann spinor.
  In the present context this is
equivalent to acting
on the dilaton one-point function with the broken supersymmetries which are generated
 by $\eta^A Q_A$ and $\rho_A \bar Q^A$.
 The  parameters $\eta$ and $\rho$ are identified with the components
 of $\theta$.  As described more fully in
\refs{\greenguteffects},
these one-point functions are contained in the  type IIB scalar on-shell superfield
 of \refs{\howest}.  This is a field
$\tPhi(x,\theta,\bar\theta)$ that satisfies the chirality condition
$\bar D\tPhi =0$ as well as the `on-shell' condition
$D^4 \tPhi = \bar D^4\bar \tPhi$ (where $D$ is the
type IIB covariant derivative)  that restrict its components to satisfy the equations of
motion.  This field has the expansion
\eqn\psiexpan{\tPhi(\theta)
= a + 2\theta \gamma^0
{\hat\Lambda}+ {1\over 24} \theta\gamma^0 \gamma^{\mu\nu\rho}
\theta \, G_{\mu\nu\rho} +\dots -{i\over 48} \theta\gamma^0 \gamma^{\rho\mu\nu}
\theta \,  \theta \gamma^0 \gamma_\rho^{\ \lambda\tau}
\theta \, R_{\mu\lambda\nu\tau}  + \dots,}
which terminates after the eighth power of $\theta$. In this
expression $\gamma^\mu$ ($\mu =0, 1,\dots,9$) are the ten-dimensional
(euclidean)
gamma matrices and  the
complex fluctuation of the  scalar field
is denoted by $a$, which is defined by
\eqn\adeff{\eqalign{\tau &= \chi+ {i\over g} + {1\over g}(\hat C^{(0)}
-i \hat\phi)\cr
& = \tau_0 + {1\over g}a,\cr}}
(where the constant background scalar field is now denoted by
$\tau_0 = \chi + ig^{-1}$ and the hats denote fluctuations of fields)
and transforms with $SL(2,Z)$ weight $(2,-2)$.
The complex dilatino, \foot{The dilatino is here represented by
the symbol $\hat \Lambda$ to avoid confusion with the gaugino of
the open-string sector.}
${\hat\Lambda}$, transforms with weight $({3/2},-{3/2})$,
while  $G_{\mu\nu\rho}$ is
the field strength of a complex linear combination of \NSNS\ and
\RR\ antisymmetric tensor potentials that transforms with weight
$(1,-1)$.
The complex gravitino of weight $({3/2},-{3/2})$
occurs as the coefficient of the
term of order $\theta^3$ while  the $SL(2,Z)$-invariant
 scalar Riemann curvature occurs as the coefficient of the $\theta^4$
term (together with
 $\partial F_5$, where $F_5$ is the self-dual \RR\ field strength).
The higher powers of
  $\theta$ have coefficients that are derivatives acting on the complex
conjugate fields.

We here want to include a D3-brane in the $0,1,2,3$ directions.  The
sixteen supersymmetries broken by the D-instanton  are now distinguished by the fact that
the $\rho$ supersymmetries
are also broken by the D3-brane while the $\eta$ supersymmetries are unbroken by the
D3-brane.
 More precisely, the extra condition on the ten-dimensional
chiral spinor  $\theta$  that corresponds to  the supersymmetries that are
 preserved on the D3-brane is,
\eqn\spina{{\cal P}_+\; \theta=0  ,}
where ${\cal P_{\pm}}= (1\pm\gamma^{0123})/2$ is a projection operator.
The solution of this equation has eight independent
components that are associated with the $\eta$ supersymmetries that act as shifts on
$M'$.
Likewise, the remaining eight components satisfy ${\cal P}_-\theta
=0$. These
are to be identified with the with the $\rho$ supersymmetries that act as shifts on
$\lambda_A^\dal$ and $\bar\Lambda_A^\dal$.
The zero modes resulting from the breaking of the bulk supersymmetries
 are therefore obtained by
expressing the superfield $\tilde \Phi$ as a power series in  $M'$ and
$\lambda$.

  For example, one contribution to
the coupling of  the graviton to the disk is obtained  by acting four times
 with the broken $\eta$ supersymmetries on the dilaton one-point
function, which  picks out the fourth power of $\theta$ in the superfield
$\tPhi$,
\eqn\gravcoup{\langle h\rangle_4= h_{\mu\nu}k_\lambda k_\rho
  \theta\gamma^0 \gamma_\tau^{\ \mu\lambda}\theta\,  \theta\gamma^0\gamma^{\tau
\nu\rho}\theta.}
In the  `minimal' case that we considered earlier, in which
 all the $\mu$ and $\lambda$ fermions are integrated out
 by bringing down factors of $\lambda \mu\bar w$ and $\lambda \bar\mu
 w$ from $e^{-S_{-1}}$ , the  non-zero graviton D-instanton induced
  two-point function is  given
 purely  by the
 integration over the  eight components of $\theta$ that correspond to
 the $M'$ coordinates,
\eqn\gravintre{  \eqalign{ A^{inst}_{R^2}&={\cal C} \int {d^3W^c\over
 W_0} d^6 \chi
d^8 M'\,   \langle h_1\rangle_{4}\,
\langle h_2\rangle_{4}\cr
& = {\cal C}\int {d^3
W^c\over W_0} d^6 \chi d^8 M'\,  e^{-S_{-1}}
   h^{(1)}_{\mu_1\nu_1}k^1_{\lambda_1} k^1_{\rho_1}
  \theta\gamma^0\gamma^{\tau_1 \mu_1\lambda_1}\theta\,
\theta\gamma^0\gamma^{\tau_1
\nu_1\rho_1}\theta\cr
&\qquad  h^{(2)}_{\mu_2\nu_2}k^2_{\lambda_2} k^2_{\rho_2}
  \theta\gamma^0\gamma^{\tau_2 \mu_2\lambda_2}\theta\,
\theta\gamma^0\gamma^{\tau_2
\nu_2\rho_2}\theta.}}
In order to evaluate the fermionic integrals
 and show that this expression  is proportional  to the kinematic factor
\kinematicfac\ we need to project $\theta$ onto its
$\lambda$ and $M'$ components.
It is convenient to use a special frame in which $SO(10)$ is broken to
$SO(2)\times SO(8)$, where the $SO(2)$ refers to rotations in the $\mu=0,9$
plane and the momenta
 and polarizations of the graviton wave functions are
chosen to be independent of the $\mu=0,9$ directions.
The
D$(-1)$/D3 background  breaks the $SO(8)$ to
$SO(5)\times SO(3)$ (the two factors being
associated with the directions $4,5,6,7,8$ and $1,2,3$, respectively),
which is also a subgroup of the full $SO(6) \times SO(4)$ symmetry
group.   The chiral $SO(6)\times SO(4)$ bispinor
$M'$ can then be identified with a chiral $SO(8)$ spinor.
More explicitly, the sixteen-component spinor  $\theta$ decomposes
under $SO(8)$ into
two  spinors of opposite chiralities,  $\theta=M_c + M_s$,
where $M_c=
(1+i\gamma^{09})\theta/2$ and $M_s=(1-i\gamma^{09}) \theta/2$.
Since $\theta$ satisfies the condition
\spina\ it is possible to write $\theta= 2 {\cal P}_-\,\eta_s$ where
$\eta_s$ is a
${\bf 8_s}$ spinor.  It follows that  $M_s= \eta_s$ and
$M_c=\hat\gamma^{123}\, \eta_s$.
It is then straightforward to show
(using an explicit representation of $SO(10)$ gamma
matrices  in terms of
$SO(8)$ gamma matrices) that
\eqn\lcdecomb{\eqalign{h_{\mu\nu} k_\rho k_\lambda\,
\theta\gamma^0\gamma^{\tau\mu\lambda}\theta\,
\theta\gamma^0\gamma^{\tau\nu\rho}\theta &\to
h_{\mu\nu}k_\lambda k_\rho\;
M_c\hat\gamma^{\mu\lambda}M_c\;M_s\hat \gamma^{\nu\rho}M_s  \cr
& = h_{\mu\nu}k_\lambda k_\rho\; \eta_s \hat\gamma^{\mu\lambda}\eta_s
\; \eta_s  \hat\gamma^{123}\hat \gamma^{\nu\rho}\hat
\gamma^{123}\eta_s\cr
&= h_{\mu\bar\nu}k_{\lambda} k_{\bar\rho}\;D^{\bar\rho}_\rho
D^{\bar\nu}_\nu \;\eta_s
\hat\gamma^{\mu\lambda}\eta_s \;\eta_s\hat \gamma^{\nu\rho}\eta_s.\cr}}
On the right hand side of this equation
 $\hat \gamma$ indicates a $8\times 8$  $SO(8)$ gamma matrix, and
$M_s$, $M_c$ and $\eta_s$  are now eight-component $SO(8)$ spinors
(and the vector indices are in the range
$\mu,\nu,\lambda,\rho=1,\cdots,8$) and
use has been made of $SO(8)$ Fierz relations.

Inserting \lcdecomb\ into \gravintre\ and identifying  $\eta_s$  with
$M'$ leads once again to the standard integration
 over the product of  components of a fermionic $SO(8)$
spinor (as in \mints) that results in contractions with the tensor,
$t_8$.  The resulting D-instanton induced two-graviton amplitude
is
\eqn\graviinstind{A^{inst}_{R^2}={\cal C}\;(\alpha')^2\;e^{2\pi i \tau}\;
 t_8^{\mu_1\nu_1\rho_1\lambda_1 \mu_2\nu_2\rho_2\lambda_2 } \;D^{\bar
 \rho_1}_{\rho_1}
 D^{\bar \lambda_1}_{\lambda_1} \;D^{\bar \rho_2}_{\rho_2} D^{\bar
\lambda_2}_{\lambda_2}\;
h^{(1)}_{\mu_1 \bar\rho_1}k^1_{\nu_1}
k^1_{\bar \lambda_1}h^{(2)}_{\mu_2 \bar\rho_2}k^2_{\nu_2}
k^2_{\bar \lambda_2} +c.c.}
(absorbing a combination of constants into the definition of
${\cal C}$).
After covariantizing this  expression  it is
proportional to the $(\alpha')^2$ term of  the tree level result \kinematicfacb.

A similar discussion leads to the other
 instanton-induced interactions of bulk supergravity
fields together with the D3-brane supersymmetric Maxwell fields that
 break 24 of
the original 32 supersymmetries (i.e., are in the
`minimal' sector).  More generally, the presence of closed string sources
leads to sectors in which more supersymmetry is broken.  As we saw, this arises
from terms in which some, or all, of the factors of $\mu$ and $\bar\mu$ are
soaked up by $\mu\bar \mu$ bilinears in the expansion of $e^{S_{-1}}$.  This is
the situation in which some, or all, powers of $\lambda$ are taken from the
closed-string superfield $\tilde \Phi(M',\lambda)$.  The sector in
 which all the
$\lambda$'s are soaked up by the closed-string sources is the one in which all
32 supersymmetries are broken.  In this case the D-instanton carries
 the same set
of zero modes as the isolated bulk D-instanton and behaves independently
of the D3-brane.

\newsec{Discussion}
 We have discussed  some low energy
aspects of terms of order ${\alpha'}^4$ in the low energy
expansion of the world-volume action of a D3-brane.  The effects
of a D-instanton on such higher-derivative interactions were
obtained by explicit integration over the collective coordinates.
This resulted in  a  variety of interactions between the
world-volume fields  with a kinematic structure that matches the
tree-level and one-loop interactions  that arise in  the low
energy limit of perturbative
 open string theory.   This motivated conjectures for the
exact non-perturbative $SL(2,Z)$-invariant interactions.
Furthermore, these combine with terms quadratic in the tangent and
normal curvatures induced on the world-volume where the
dependence on the second fundamental form is built in by
interactions of the open-string scalar field, $\varphi^c$.

The lowest order contributions to these instanton amplitudes are
associated with
 disk diagrams with insertions of an even number of twist operators
 (which are the vertex operators for the massless ND modes) that change
 the  boundary condition from Neumann to Dirichlet.  The
 integration over the bosonic moduli $W^c$ and $\chi^a$ is gaussian
 and peaks at the origin where the D-instanton coincides with the
 D3-brane. Such effects of D-instantons inside D-branes have appeared in other
contexts, for example there are instanton corrections to
$\tr(F^4), (tr(F^2))^2$ and related terms in the world-volume of
D7-branes. Such terms can be calculated in special situations
\refs{\lercheetal,\gutperlea} using heterotic/type I duality and
the modular functions appear in that case are again related to
\modfun.

In our discussion we have not taken the   limit of  \refs{\malda},
in which gravity  decouples from the excitations on the brane and
which is at the heart of the AdS/CFT correspondence
\refs{\review}. The effects we have found are therefore different
from the instanton effects found in \refs{\doreythree,\mbgitaly}
which relate the induced   D-instanton interactions  to
correlators in the CFT. In the AdS/CFT decoupling limit (the
near-horizon limit of the classical D3-brane geometry)  the
D-instanton exists on the Higgs branch of the moduli space and the
`center of mass' U(1) gauge field
 discussed in this paper decouples.

In the gauge field  theory description of   $Dp/Dp+4$ systems the
singularities associated with abelian instantons or   `small'
non-abelian instantons are often problematic.  The non-zero length
scale that appears in  string theory in a trivial background
resolves these singularities which reappear in the low energy
decoupling limit that reduces to the  field theory in the absence
of background bulk fields. However, it is well-known that such
small instanton singularities are also smoothed out by the
presence of a background antisymmetric tensor field, $B_{\mu\nu}$
\refs{\aharonyseib,\nekrasovschwarz}.  In this case the decoupling limit is
non-trivial even in the abelian case and is known to correspond to
a non-commutative version of the gauge theory
\refs{\nekrasovschwarz,\seibergwitten}. The background field is
introduced by coupling the $B_{mn}$ closed-string vertex operator
to the disk with two twist operators on the boundary.
  For constant $B$ this is a total derivative and
the result is  a term in the action proportional to $\langle B \rangle
= B_{mn} W^{mn}$ which combines with the $F_{mn}$ term
\simpsource\ of section 4 to give the dependence on $2\pi \alpha'
{\cal F} =2\pi \alpha' F + B$ that is required in order to
maintain the antisymmetric tensor  gauge invariance. The action of
the eight broken supersymmetries on $\langle B \rangle$ generates a
 supermultiplet of one-point functions that combine with the
 open-string supermultiplet $\Phi_{mn}$ of section 4.  This
 description is valid for weak $B_{mn}$.    A more complete
analysis of this situation would take into account the fact that
the background field induces a shift in the Neumann boundary
conditions on the disk (turning them into Dirichlet conditions in
the large $B$ limit).

\bigskip
\noindent{\bf Acknowledgments}
\medskip
 We are grateful to Juan Maldacena, Boris Pioline  and  Wati
Taylor for useful conversations. We are also grateful for the
hospitality of the CERN
Theory Division  where this work was started.
 The work of MG is supported in part by the David and Lucile Packard
Foundation.

\appendix{A}{Conventions}

We are interested in properties of
$SO(4)$ and $SO(6)$ spinors.  For the $SO(4) = SU(2)_L \otimes SU(2)_R$
case we will use the conventions of Wess and Bagger
\wessbagger\ with the choice of representation of $4\times 4$ gamma
matrices,
\eqn\gammfour{\gamma^n =\pmatrix{0 & \sigma^n\cr
         \bar\sigma^n & 0 \cr}, \qquad \gamma_5 = \pmatrix{1 & 0 \cr
                         0 & -1 \cr}, }
where the four  matrices
$\sigma^n_{\alpha \dot \beta}$'s are   $\sigma^0= I$ $(n=0)$
and $ i\sigma^i$ ($n=1,2,3$) where   $\sigma^i$ are the  Pauli matrices
(and the factor of $i$ comes from the continuation to euclidean
signature).  A bar denotes the reversed assignment of spinor
chiralities so that the indices on $\bar \sigma^n$ are  $\bar
\sigma^n_{\dot \alpha
\beta}$.  Spinor indices are raised by means of the antisymmetric tensors,
$\epsilon^{\alpha\beta}$ and $\epsilon^{\dot\alpha,\dot\beta}$.
The $SO(4)$ group generators can be written as
\eqn\gens{\gamma^{mn} = {1\over 4}\left[\gamma^m,\gamma^n\right] =
i\pmatrix{\sigma^{mn} & 0 \cr
           0& \bar \sigma^{mn}\cr},}
where,
\eqn\lorgen{\sigma^{mn}_{\alpha\beta} =
{1\over 2} \left(\sigma^m \bar \sigma^n
- \sigma^n \bar \sigma^m\right)_{\alpha\beta} ,
\qquad
\bar\sigma^{mn}_{\dot\alpha\dot\beta} =
{1\over 2} \left(\bar \sigma^m  \sigma^n
-\bar  \sigma^n  \sigma^m\right)_{\dot\alpha\dot\beta}.}
These  are the generators of
Lorentz transformations on chiral spinors satisfying,
\eqn\sdcon{\epsilon^{mnpq}\sigma_{pq}=-2 \sigma^{mn},\qquad
\epsilon^{mnpq}\bar\sigma_{pq}=2 \bar\sigma^{mn} }
so they project on self-dual and anti  self-dual tensors respectively.
With this definition the coupling of  $F^\pm$ is given by
$\sigma^{mn}F^-_{mn}$ and $\bar\sigma^{mn}F^+_{mn}$.

The 'tHooft symbol $\eta^c_{mn}$ maps the self-dual $SO(4)=SU(2)_L \times
SU(2)_R$
tensor into the adjoint of one $SU(2)$ subgroup, while $\bar\eta^c_{mn}$
performs the mapping on the conjugate representations, so that,
\eqn\etadef{\sigma^{mn}_{\alpha\beta} = \eta^c_{mn}\,
\tau^c_{\alpha\beta},\qquad
\bar\sigma^{mn}_{\dot\alpha\dot\beta} = \bar\eta^c_{mn}\,
\bar\tau^c_{\dot\alpha
\dot\beta}.}
These symbols may be explicitly represented in the form,
\eqn\etaexp{\eqalign{ \eta^c_{mn}& = \bar\eta^c_{mn} = \epsilon_{cmn},
\qquad
m,n\in \{1,2,3\},
\cr
\bar\eta^c_{4n}& = -\eta^c_{4n} =\delta_{cn},\cr
\eta^c_{mn}   &= - \eta^c_{nm}, \qquad \bar \eta^c_{mn}  = - \bar
\eta^c_{nm}.
\cr}}
We also note the formula for the contraction of two $\eta$ symbols,
\eqn\tautau{\delta^{c_1c_2}\eta^{c_1}_{m_1n_1}\eta^{c_2}_{m_2n_2}=
\delta^{m_1m_2}\delta^{n_1n_2}-\delta^{m_1n_2}\delta^{m_2n_1}+
\epsilon^{m_1n_1m_2n_2}.}
The $SO(6)= SU(4)$ gamma matrices are $8 \times 8$  matrices that
may be   represented by,
\eqn\gammsix{\Gamma^a = \pmatrix{0 & \Sigma^a \cr
    \bar \Sigma^a & 0\cr}, \qquad \Gamma_7 = \pmatrix{-1 & 0 \cr
          0 & 1\cr}, }
where the matrices $\Sigma^a_{AB}$ and $\bar \Sigma^{a\ AB}$ are the
Clebsch--Gordan coefficients that couple two ${\bf 4}$'s of
$SU(4)$ to a ${\bf 6}$ and  two ${\bf \bar 4}$'s to a ${\bf 6}$,
respectively.  These can also be written in terms of the 'tHooft
symbols,
\eqn\bigsig{\eqalign{
\Sigma^a_{AB} & = \eta^c_{AB}, \qquad (a=1,2,3);\qquad
\Sigma^a_{AB}  = i\bar \eta^c_{AB}, \qquad (a=4,5,6)\cr
\bar\Sigma_a^{AB} & = -\eta^c_{AB}, \qquad (a=1,2,3); \qquad
\bar \Sigma_a^{AB}  = i\bar \eta^c_{AB}, \qquad (a=4,5,6)
.\cr}}
The   $SO(6)$ generators can be represented by
\eqn\biggamab{
\Gamma^{ab} = {i\over 4} \left[\Gamma^a, \Gamma^b \right]=
i\pmatrix{\Sigma^{ab} & 0 \cr
           0& \bar \Sigma^{ab}\cr}.}

The charge conjugation operator for the $SO(4)$ group is
\eqn\conjfour{C_4 = \pmatrix{-i\sigma^2 & 0 \cr
           0 & -i\sigma^2\cr},}
while the charge conjugation operator for $SO(6)$ is
\eqn\conjsix{C_6 = \pmatrix{0& -i\cr
              -i & 0 \cr}.}

\appendix{B} {Curvature induced from embeddings}

This appendix gives a very brief summary of
some standard notation concerning embedded $(p+1)$-dimensional
 sub-manifolds (for further details  see \refs{\eisen,\koba} and the summary in
 \refs{ \bainbachasgreen}).
The embedding  coordinates $Y^\mu (x^m)$ (where
 $\mu = 0,\cdots, 9$ is the target space-time index and
   $m=0,\cdots p$ the  world-volume index) describe the position of the
 $(p+1)$-dimensional  world-volume in the target space.
 The  quantity $\partial_m Y^\mu $
   defines a local frame for  the tangent bundle while a frame for the
    the normal bundle,  $\xi^\mu _a$
 ($a=p+1,\cdots  9$), is defined by
\eqn\norm{
\xi^\mu _{a}\;\xi^\nu_{b}\; G_{\mu \nu} = \delta_{ab}\ \ \ \ {\rm
and} \ \ \ \xi^\mu _{a}\; \partial_m Y^\nu\;  G_{\mu \nu} = 0 \ .
} Both   $\partial_m Y^\mu $ and  $\xi^\mu _{a}$ transform as
vectors under target-space reparameterization. The former are
vectors of world-volume reparametrizations, while the latter
transform as vectors under local $SO(9-p)$ rotations of the normal
bundle. Correspondingly, the metric on the D-brane world-volume
can be decomposed as
\eqn\comp{
G^{\mu \nu} = \partial_m Y^\mu
 \partial_n Y^\nu\; { g}^{mn}
+ \xi^\mu _a \xi^\nu_b\; \delta^{ab}\ . } For many purposes it is
convenient to use the static gauge in which $Y^m(x^m) = x^m$ (when
$\mu = 0,1,\dots,p-1$).  In this gauge the transverse coordinates
are $Y^a$ (where $\mu = p, \dots,9$) and are interpreted as the
scalar fields $\varphi^a(x^m)$,
 that enter into the D$p$-brane action.

Any tensor can be pulled back from the target space onto the
tangent or normal bundles by contraction with the local frames.
Thus, the pull-back of the bulk (target-space) metric  is the
induced  world-volume metric,
\eqn\metric{
{ g}_{m n} = G_{\mu \nu}\;
\partial_m Y^\mu \partial_n Y^\nu
\ . } The Riemann tensor  can be pulled back in different ways,
depending on whether any of its four indices are contracted with
the tangent or normal frame.

The   target-space connection
 $\Gamma^\mu _{\nu\rho}$ can be constructed from the target-space metric while
the  world-volume connection  ${{\Gamma}_{_{T}}}^m_{n\gamma}$ can
be
 constructed in terms of the induced metric.  The connection on the normal bundle
 is a composite SO(9-p) gauge field
\eqn\gauge{
\omega^{ab}_{ m} = \xi^{\mu , [a}\; \left( G_{\mu \nu}  \partial_m
+  G_{\mu \sigma} \Gamma^{\sigma}_{ \nu\rho}
 \partial_m Y^\rho \right)\;  \xi^{\nu,  b]}\ ,
} which may be defined   by requiring
 that the normal frame be  covariantly constant.
The covariant
 derivative of the tangent frame, which is the second fundamental form, is given by
 \eqn\second{
 \Omega_{mn}^\mu = \Omega_{nm}^\mu =
  \partial_m \partial_n Y^\mu  -
  ({\Gamma}_T)^\gamma_{mn}\;
  \partial_\gamma Y^\mu + \Gamma^\mu _{\nu\rho}\;
  \partial_m  Y^\nu \partial_l Y^\rho \ ,
  }
and  is a symmetric world-volume tensor and  space-time vector.
The
 tangent-space pull-back of $\Omega$ vanishes so that
$\Omega^l_{mn}=0$, which implies that the non-zero components are
\eqn\sec{
\Omega_{mn}^a  \equiv  \Omega_{mn}^\mu \; \xi^{\nu,
 a}
G_{\mu \nu}\ .
 }
Totally geodesic embeddings are characterized by a vanishing
second fundamental form.

The Gauss-Codazzi equations express  the world-volume curvature $
R_T$,  constructed out of the affine connection $\Gamma_T$, and
the field strength,  $R_N$, of the $SO(9-p)$ gauge connection
$\omega$,  to pull-backs of the space-time  Riemann tensor
together with combinations of the second fundamental form.  These
relations are
\eqn\gaus{
({ R}_T)_{mnpq} =  {R}_{mnpq} + \delta_{ab}\left(
\Omega^a_{mp}\Omega^b_{nq}- \Omega^a_{mq}\Omega^b_{np} \right)}
and
\eqn\gausses{
(R_N)^{\ \ \ ab}_{mn} = -R^{\ \ \ ab}_{mn} +  { g}^{pq}\left(
  \Omega^a_{mp}\Omega^b_{nq}
- \Omega^b_{mp} \Omega^a_{nq} \right) \ . } Only if the D$p$-brane
world-volume is  a totally geodesic manifold (so that $\Omega =0$)
do the curvature forms  $R_T$ and $R_N$  coincide with the
pull-backs of the bulk curvature.  For embeddings in flat
space-time these
 world-volume curvatures can be
expressed entirely in terms of the second fundamental form
 $\Omega$.
\ From \second\ we see that the presence of the terms bilinear in
$\Omega$ in \gaus\ and \gausses\ translates into
 terms of quadratic and higher order in derivatives of the scalar fields.

The bulk Ricci tensor vanishes in the backgrounds of relevance to
us.  However, the tensor $\hat R_{ab} = g^{mn} R_{m a n b}$ is
nonvanishing and enters in the expression \cpeven\ where the
combination $\bar R_{ab}$ is defined by
\eqn\rbardef{\bar{R}_{ab} \equiv {\hat R}_{a b}   +
{g}^{mm'} {g}^{nn'} \Omega_{a \vert m n}\Omega_{b \vert m'n'}.}

\listrefs

\bye